\pdfoutput=1

\documentclass[11pt]{article}

\usepackage[preprint]{acl}

\usepackage{times}
\usepackage{latexsym}

\usepackage[T1]{fontenc}

\usepackage[utf8]{inputenc}

\usepackage{microtype}

\usepackage{inconsolata}

\usepackage{graphicx}

\usepackage{subcaption}

\usepackage{float}

\usepackage{pdflscape}
\usepackage{tabularx}
\usepackage{lipsum}

\usepackage{longtable}

\title{AgentMaster: A Multi-Agent Conversational Framework Using A2A and MCP Protocols for Multimodal Information Retrieval and Analysis}

\author{Callie C. Liao \\ 
  Stanford University \\
  \texttt{ccliao@stanford.edu} \\\And
  Duoduo Liao\\
  George Mason University \\
  \texttt{dliao2@gmu.edu} \\\And
  Sai Surya Gadiraju \\
  George Mason University \\
  \texttt{sgadira3@gmu.edu} \\}

\begin{document}
\maketitle

\begin{abstract}

The rise of Multi-Agent Systems (MAS) in Artificial Intelligence (AI), especially integrated with Large Language Models (LLMs), has greatly facilitated the resolution of complex tasks. However, current systems are still facing challenges of inter-agent communication, coordination, and interaction with heterogeneous tools and resources. Most recently, the Model Context Protocol (MCP) by Anthropic and Agent-to-Agent (A2A) communication protocol by Google have been introduced, and to the best of our knowledge, very few applications exist where both protocols are employed within a single MAS framework. We present a pilot study of AgentMaster, a novel modular multi-protocol MAS framework with self-implemented A2A and MCP, enabling dynamic coordination, flexible communication, and rapid development with faster iteration. Through a unified conversational interface, the system supports natural language interaction without prior technical expertise and responds to multimodal queries for tasks including information retrieval, question answering, and image analysis. The experiments are validated through both human evaluation and quantitative metrics, including BERTScore F1 (96.3\%) and LLM-as-a-Judge G-Eval (87.1\%). These results demonstrate robust automated inter-agent coordination, query decomposition, task allocation, dynamic routing, and domain-specific relevant responses. Overall, our proposed framework contributes to the potential capabilities of domain-specific, cooperative, and scalable conversational AI powered by MAS. The demonstration video is accessible via this \href{https://youtu.be/H2SzebEJGmo}{link}.

\end{abstract}

\section{Introduction}

\begin{figure}
 \centerline{
 \includegraphics[width=0.97\columnwidth]{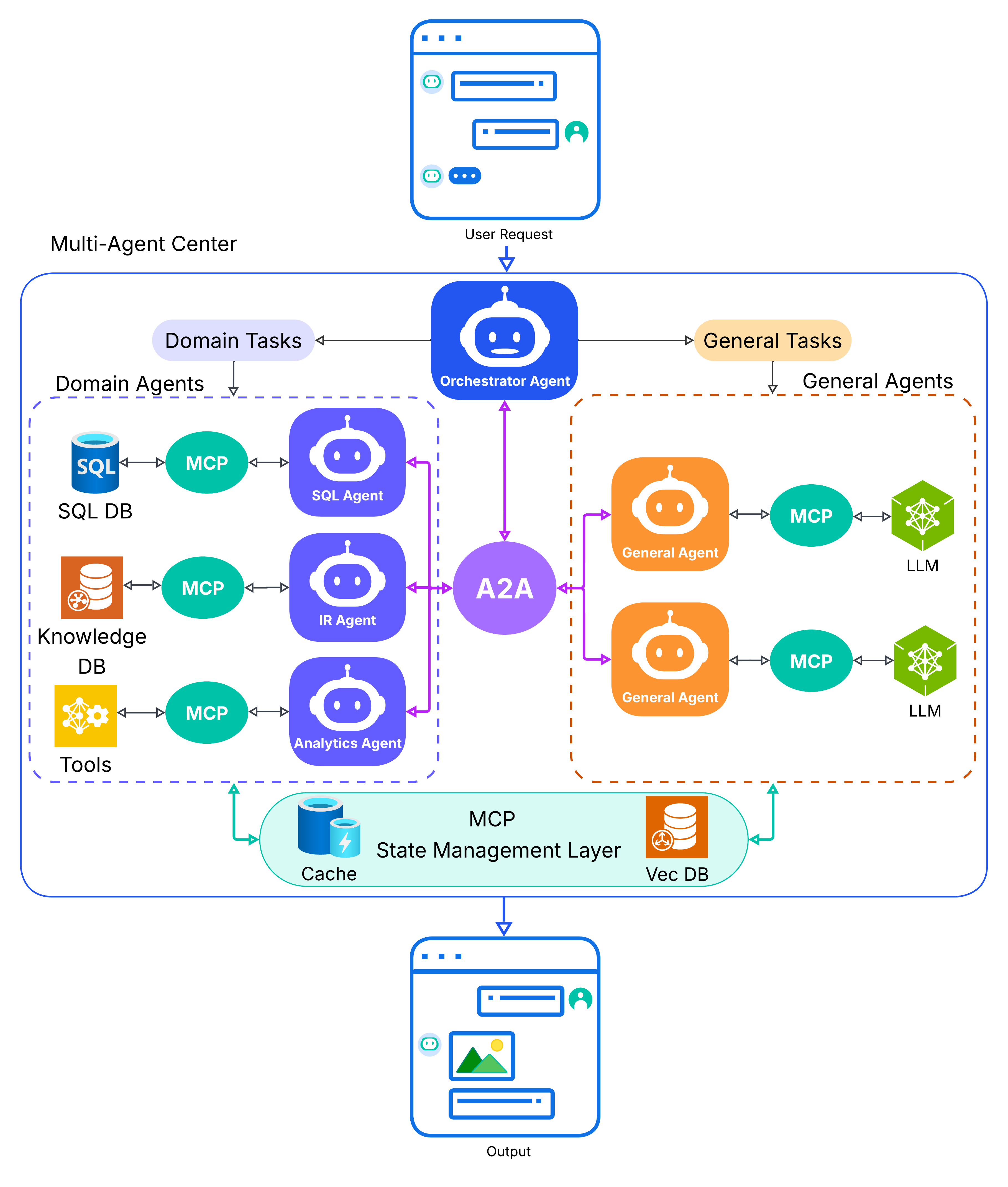}}
 \caption{The general MAS framework of AgentMaster.}
 \label{fig:sys_framework}
\end{figure}

Recent advances in artificial intelligence (AI) have increasingly focused on Multi-Agent Systems (MAS), in which multiple intelligent agents collaborate, communicate, and share contextual information to address complex tasks \citep{li2025agentorientedplanningmultiagentsystems, ACL2024_qian-tell, ICLR-yao2023reactsynergizingreasoningacting}. The integration of Large Language Models (LLMs) into MAS frameworks has significantly broadened their applicability, enabling general-purpose collaboration, natural language interaction, and open-ended reasoning \citep{hu2025compileagentautomatedrealworldrepolevel, ACL-luo-2024-RepoAgent, ACL2024_huang-queryagent, guo2024largelanguagemodelbased}. This makes LLM-based MAS particularly well-suited for dynamic, unstructured tasks such as multimodal data analysis, research automation, and intelligent assistance \citep{dong2024villageragentgraphbasedmultiagentframework, ACL2025_islam-mapcoder, lin2025factauditadaptivemultiagentframework}. By distributing intelligence across agents, LLM-based MAS offer a promising approach to overcoming the limitations of standalone LLMs \citep{geminiteam2025geminifamilyhighlycapable, openai2024gpt4technicalreport, LLaMA2023}.

Despite their potential, current LLM-based MAS face critical challenges that limit their scalability, robustness, and effectiveness. These challenges span technical, architectural, and practical dimensions, including agent coordination, communication, interaction with heterogeneous tools and data sources, knowledge representation and reasoning, modularity, and integration of domain-specific expertise \citep{du2025surveycontextawaremultiagentsystems, shen2023hugginggptsolvingaitasks}. Especially, in domain-specific contexts where specialized agents are increasingly essential \citep{NIPS2025_FinCon_MAS, ACL-mathur-2024-MATSA, InfoTechBot-BigData2024}, these systems often require substantial domain-specific knowledge and the capability to process diverse data modalities, posing additional challenges for effective automation and coordination \citep{NIPS2025_FinCon_MAS, bio_agents_2025, haase2025staticresponsesmultiagentllm, Amazon_2025_supplychain_MAS}.

Most recently, two new open standards, Anthropic’s Model Context Protocol (MCP) \citep{Anthropic_MCP2024} and Agent-to-Agent (A2A) communication protocol introduced by Google \citep{Google_blog_A2A2025}, aim to address these challenges. MCP, announced in May 2024, streamlines the process by providing a standardized interface for accessing various tools and resources, enhancing the modularity, interoperability, and statefulness of multi-agent and tool-augmented systems. A2A, announced in May 2025, complements MCP by facilitating structured inter-agent communication, which allows multiple AI agents to exchange messages, distribute subtasks, and build shared understanding to solve problems collectively. Both A2A and MCP can be developed using existing SDKs or fully implemented by users as needed. These protocols offer a systematic alternative to the fragmented, ad hoc integration approaches common in current MAS implementations \citep{A2A-MCP_jeong2025studymcpxa2a, A2A-MCP_yang2025surveyaiagentprotocols}.

Existing LLM-based multi-agent systems that do not incorporate A2A or MCP often suffer from static coordination, limited memory, and rigid communication mechanisms. By leveraging these emerging standards, systems can support structured inter-agent communication, maintain shared contextual understanding, and seamlessly interface with external tools, developing more capable, scalable, and cooperative AI systems \citep{A2A-MCP_yang2025surveyaiagentprotocols, A2A-MCP_ehtesham2025surveyagentinteroperabilityprotocols}.

To date, both industry and academia have conducted limited research on the application of A2A and MCP within LLM-based MAS. While a few research efforts have explored the independent use of A2A \citep{A2A_habler2025buildingSecureAgenticAI} and MCP \citep{MCP_krishnan2025advancingmultiagentsystemsmodel, MCP_qiu2025agentdistilltrainingfreeagentdistillation, MCP_sarkar2025surveyllmagentcommunication}, there are, to the best of our knowledge, very few applications in which both protocols have been jointly employed within a single MAS framework. 

\begin{figure*}
    \centering
    \includegraphics[width=\textwidth]{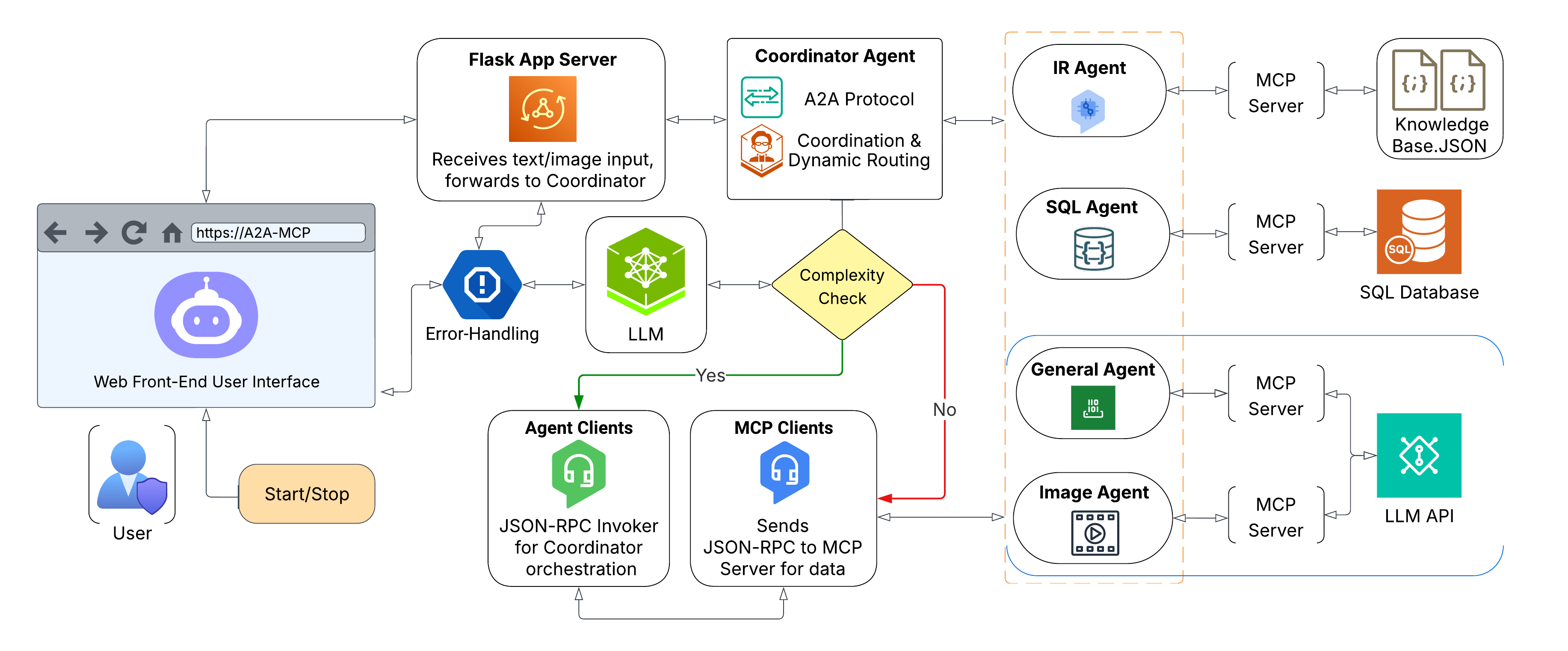}
    \caption{The system architecture of the case study.}
    \label{fig:sys_arch}
\end{figure*}

To address these gaps, this paper introduces AgentMaster, a novel modular multi-protocol MAS framework that integrates A2A protocol and MCP. AgentMaster decomposes user queries into specialized workflows executed by dedicated agents, coordinated through A2A and supported by a centralized MCP backend for tool and context management. Users interact with the system through a unified conversational interface, enabling natural language interaction without prior technical expertise. The framework supports automated complex task decomposition, dynamic routing, and agent-to-agent orchestration. By isolating agents and provisioning separate API keys, the system can manage resource utilization and enforce the separation of concerns between components. 

A fully functional prototype through self-developed A2A and MCP demonstrates AgentMaster’s capabilities in domain-specific multimodal tasks, including information retrieval, image analysis, database querying, question answering, and content summarization. The system is deployed both locally and on Amazon Web Services (AWS) as a set of Flask-based microservices, and exhibits consistent performance across varied task types in a pilot study.

Our main contributions are as follows:
\begin{itemize}
    \item This paper introduces AgentMaster, a modular multi-agent MAS framework that integrates Anthropic's MCP and Google's A2A protocol to enable flexible inter-agent communication, intelligent coordination, and retrieval-augmented generation.
    \item A unified system architecture is designed to support automated query decomposition, task allocation, dynamic routing, and orchestration across specialized retrieval agents and multimodal data sources.
    \item The pilot study explores the implementation of self-developed A2A and MCP protocols specifically designed for AgentMaster without relying on existing libraries such as Google’s A2A SDK.
    \item Comprehensive evaluation is conducted using G-Eval, BERTScore, and related metrics to validate correctness, completeness, and semantic fidelity across diverse query types.
\end{itemize}

\section{The General System Framework}

Figure \ref{fig:sys_framework} illustrates the general multi-protocol MAS architecture of the AgentMaster. The framework comprises four core components: a unified conversational interface, a multi-agent center, multi-agent AI protocols, and a state management layer.

\subsection{Unified Conversational Interface}
The unified conversational interface resembles a chatbot, receiving user input in various multimodal formats, including text, charts, images, and audio, and generating corresponding output in modalities such as text, images, and structured data tables. 

\subsection{Multi-Agent Center}

The Multi-Agent Center consists of three hierarchical layers of agents: the orchestrator agent, domain agents, and general agents. At the top of the hierarchy, the orchestrator agent is responsible for decomposing tasks and coordinating execution across agents. Domain agents specialize in specific functionalities and may be either LLM-based or non-LLM-based. General agents operate independently, each paired with a dedicated LLM to handle general-purpose reasoning tasks. All agents communicate through the A2A protocol, which enables structured, language-based message exchange. Additionally, each agent is integrated with the MCP protocol, which standardizes interactions with external tools, APIs, and contextual resources.

\begin{figure*}[t!]
    \centering
    \begin{subfigure}[t]{\textwidth}
        \centering
        \includegraphics[height=3in]{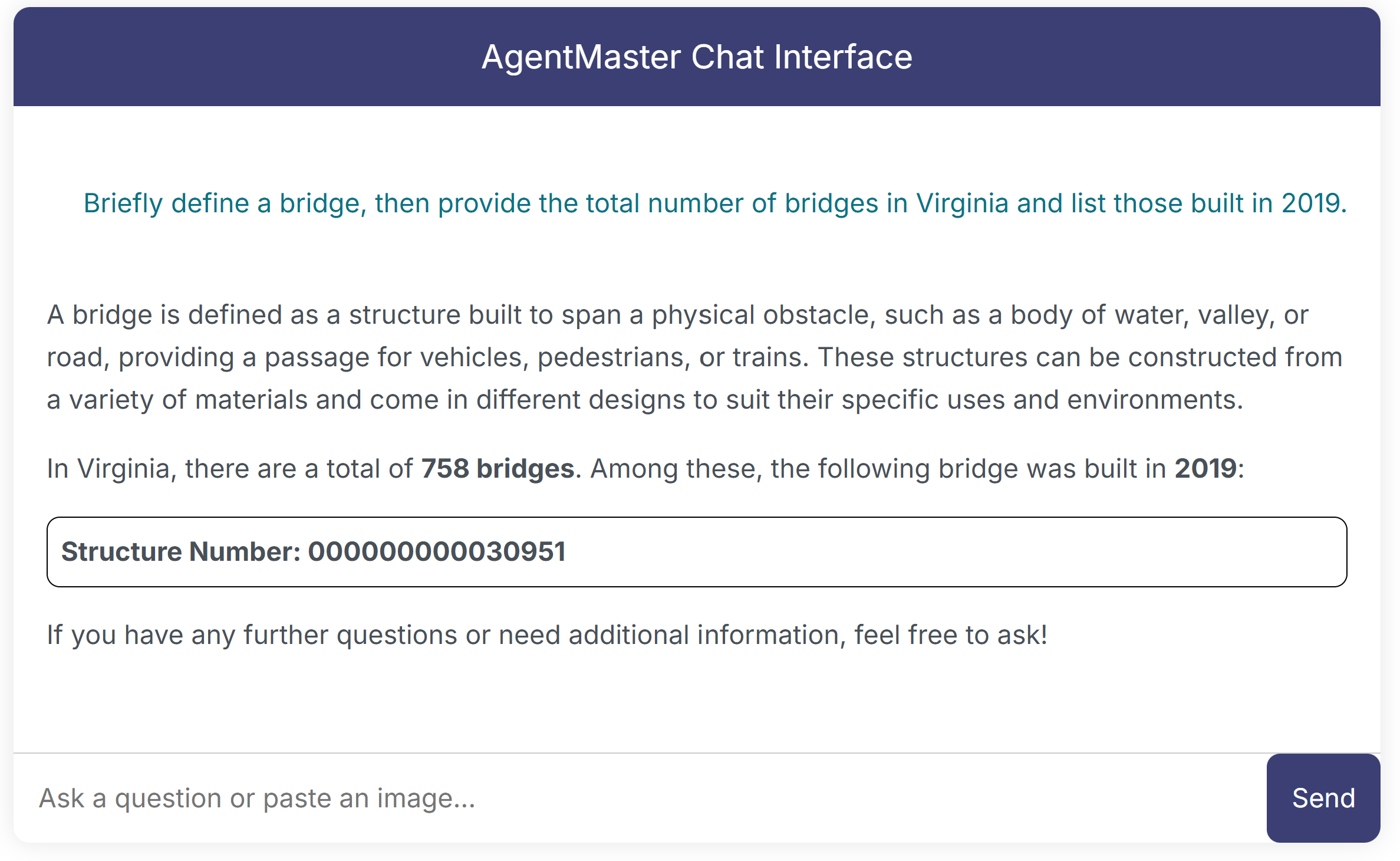}
        \caption{Frontend example.}
        \label{fig:demo_frontend}
    \end{subfigure}%
    
    \begin{subfigure}[t]{\textwidth}
        \centering
        \includegraphics[height=1.8in]{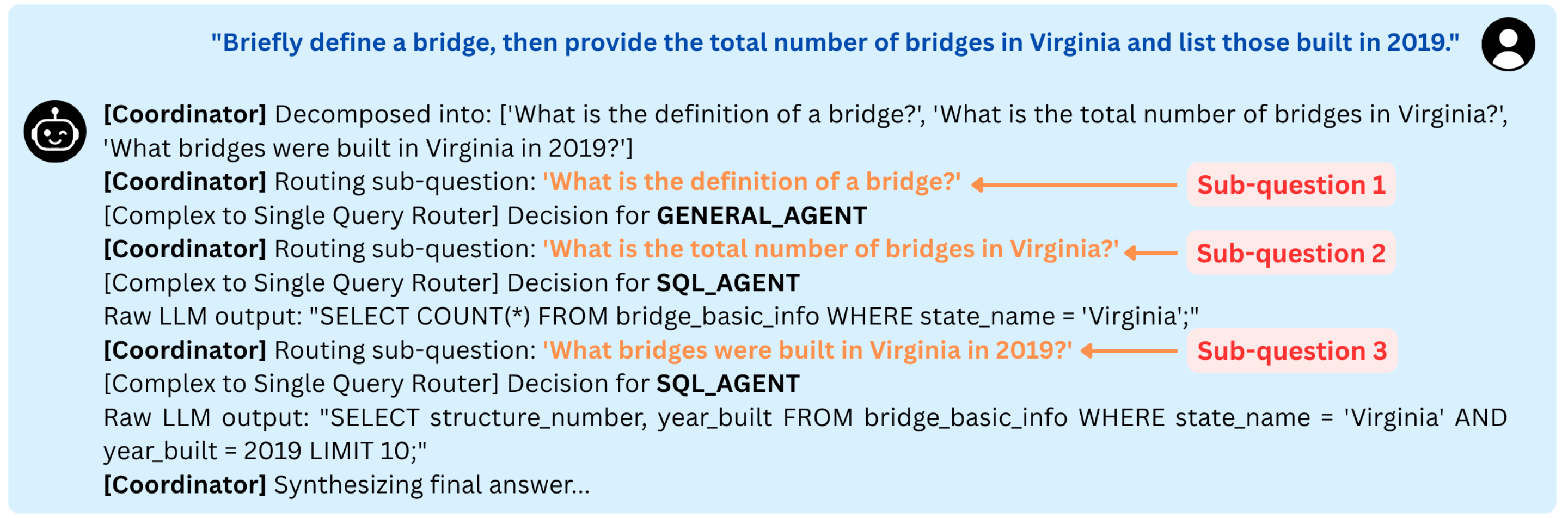}
        \caption{Backend example.}
        \label{fig:demo_backend}
    \end{subfigure}

    \begin{subfigure}[t]{\textwidth}
        \centering
        \includegraphics[height=2.5in]{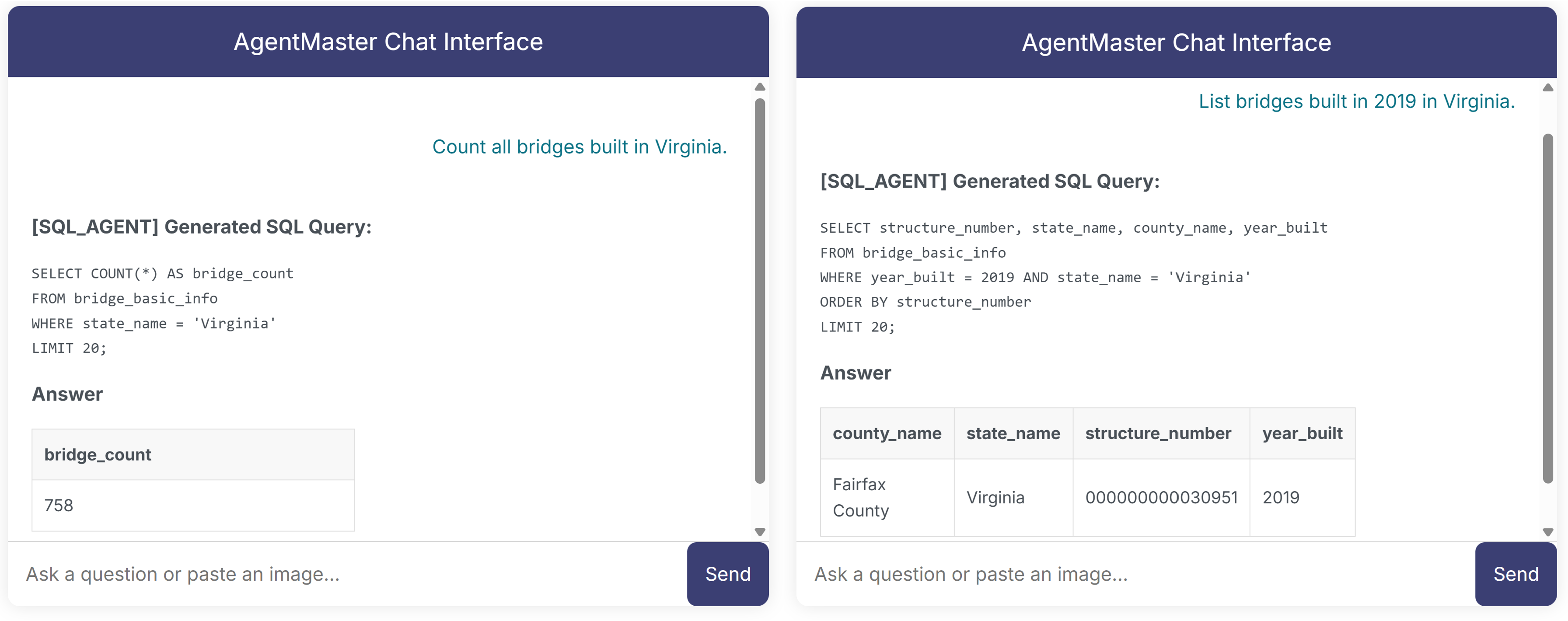}
        \caption{Verification of the generated responses to the complex query. }
        \label{fig:demo_check}
    \end{subfigure}%
    
    \caption{AgentMaster demonstration example and verification.}
    \label{fig:agent_master_demo}
\end{figure*}

\subsubsection{Orchestrator Agent} 
The orchestrator agent serves as the central coordinator, identifying available tasks and delegating them to appropriate agents based on their capabilities. To optimize efficiency and accuracy, it may further decompose complex user requests into subtasks for parallel or sequential execution across agents. As a pivotal hub, it not only translates high-level user goals into manageable tasks aligned with agent capabilities, but also facilitates inter-agent communication, handles error management across protocols, and synthesizes outputs into a coherent, unified response.

\subsubsection{General Agents} The general agent is designed to handle broad tasks that do not require access to domain-specific datasets. The orchestrator agent determines whether to delegate a task to a general agent or a domain agent, selecting the most appropriate agent based on the nature and complexity of the task.

\subsubsection{Domain Agents} Domain agents are specialized for specific domains and designed to interface with domain-relevant functions, datasets, and tools. Each domain agent may internally manage sub-agents to further decompose and process tasks in a modular fashion. These agents communicate not only with each other, but also with general agents, enabling collaborative task execution across domains. 

In AgentMaster, domain agents are designed to specialize in common domain-specific tasks, including Structured Query Language (SQL) querying, Information Retrieval (IR), and multimodal data analytics. The framework is extensible, allowing for the integration of additional agents as required to support diverse application needs.

\subsection{Multi-Agent AI Protocols}

AgentMaster employs A2A for structured communication between agents, enabling coordination, delegation, and orchestration through standardized JSON-based message exchange. MCP complements this by providing a unified interface for tool access, long-term memory, and context management, enhancing modularity, interoperability, and statefulness in LLM-based agents.

Depending on the application and requirements, the framework leverages the A2A protocol via Google's A2A SDK \citep{Google_blog_A2A2025}, or fully implements it as needed. MCP is developed in a similar manner.
 
\subsection{State Management Layer}

The State Management Layer in AgentMaster leverages vector databases and context caches to maintain the MCP state, enabling agents to be context-aware and memory-augmented for efficient handling of multistep, user-specific, and domain-specific tasks. This layer utilizes the vector database to provide persistent semantic memory for retrieving relevant past interactions and documents, while the context cache offers fast, temporary storage for session data and intermediate results during active workflows.

\section{System Architecture of the Case Study}

Figure~\ref{fig:sys_arch} illustrates the architecture of a conversational MAS, an example implementation of the AgentMaster framework for multimodal information retrieval and analysis. The system integrates modular components to enable robust, retrieval-augmented question answering through dynamic agent orchestration.

The architecture comprises a web-based user interface, a Flask server acting as the main entry point, a Coordinator agent (i.e., the Orchestrator agent) implementing the A2A protocol, and multiple specialized retrieval agents (i.e., domain agents). User queries are submitted via the chatbot front end and processed asynchronously by the backend components.

\subsection{Coordinator Agent and Complexity Assessment}

The Coordinator agent is responsible for query analysis, routing, and orchestration \citep{zhang2025agentorchestrahierarchicalmultiagentframework}. A key function is the complexity assessment module, which determines whether a query requires multi-agent collaboration or can be handled by a single retrieval agent. For simple queries, the Coordinator dispatches requests directly to an appropriate MCP client. In contrast, complex queries trigger agent clients that dynamically coordinate multiple retrieval workflows.

\subsection{Agent Clients and MCP Clients}

Agent clients serve as JSON-RPC invokers for orchestrating distributed workflows among retrieval agents. MCP clients manage communication with retrieval backends, dispatching JSON-RPC requests to MCP servers that encapsulate domain-specific retrieval logic \citep{kumar2025mcpguardiansecurityfirstlayer}. This division enables the system to support compositional retrieval and fallback handling without manual routing configuration.

\subsection{Retrieval Agents}

The system incorporates four primary specialized agents: 
(i) an IR agent that retrieves unstructured content from knowledge bases; 
(ii) a SQL agent that generates and executes SQL queries over relational databases; 
(iii) an image agent that processes image inputs through external vision APIs; 
and (iv) a general agent that handles open-domain queries and fallback cases. 
Each agent exposes an MCP server endpoint for standardized invocation.

\subsection{LLM Integration and Error Handling}

The architecture integrates a local or external LLM for language generation, reasoning, and summarization. The LLM module aggregates partial outputs returned by retrieval agents and formulates the final response. The Flask server and Coordinator agent include error-handling mechanisms that detect and recover from failures in retrieval workflows and model inference
\citep{williams2025multi}.

\subsection{End-to-End Workflow}

End-to-end query resolution proceeds as follows. The user submits a text or image query via the front end. The Flask server forwards the request to the Coordinator agent, which performs complexity assessment and routes the query to the appropriate retrieval pathway. Specialized retrieval agents return results via MCP clients. The LLM module synthesizes the final output, which is delivered to the user interface for presentation.

\subsection{Design Considerations}

The A2A-MCP design emphasizes modularity, extensibility, and reproducibility. New retrieval agents can be integrated without modifying the orchestration logic. The standardized JSON-RPC interfaces facilitate consistent communication across agents \citep{zhang2025agentorchestrahierarchicalmultiagentframework}. This architecture provides a flexible foundation for retrieval-augmented conversational systems and supports future research into multi-agent LLM collaboration.

\section{Experimental Results and Evaluation}

In this case study, the AgentMaster system is deployed locally as well as on AWS to facilitate internet access. Each agent leverages OpenAI’s GPT-4o mini model. Three domain agents, derived from our prior research, focus on SQL \citep{infoTechSQLBot-ACL2025}, IR \citep{InfoTechBot-BigData2024, infoTechSQLBot-ACL2025}, and image analysis \citep{Liao-ImageCaption-BigData2024}, utilizing the Federal Highway Administration (FHWA) public datasets  \citep{fhwa2024portal}.

Experiments  were conducted to evaluate both individual agents and agent-to-agent collaborations using simple and complex queries. Multiple evaluation metrics are employed to assess the multi-agent system, including agentic metrics, LLM-as-a-Judge \citep{NEURIPS2023_LLM_as_a_Judge}, and human evaluation. Agentic metrics assess autonomy and effectiveness of AI agents in complex tasks. LLM-as-a-Judge uses a large language model to evaluate outputs of another LLM for correctness, relevance, and coherence. Human evaluation remains the gold standard for validating these assessments in this pilot study.

\subsection{Individual Agent Evaluation}
Three domain agents (SQL, IR, and Image) were previously evaluated independently in our past research and demonstrated high reliability and accuracy \citep{InfoTechBot-BigData2024, infoTechSQLBot-ACL2025, Liao-ImageCaption-BigData2024}. Additionally, due to the robustness of the GPT model, individual queries or single tasks have consistently yielded correct results in our testing. However, there are occasional instances of misclassifying single queries as complex queries for query decomposition, resulting in incorrect responses.  

\subsection{Complex Task Evaluation}

To evaluate the quality and accuracy of AgentMaster's responses, sub-questions decomposed from complex queries were individually submitted to AgentMaster. The outputs generated for these simpler sub-questions were then compared to the corresponding segments within AgentMaster’s responses to the overall complex queries. Since the sub-questions are simple queries, it would not require mutli-agent collaboration and thus can serve as a verification method for AgentMaster's output. 

Figure \ref{fig:agent_master_demo} presents the front-end and back-end of the demonstration, as well as the verification of AgentMaster's generated response. As shown in Figure \ref{fig:demo_frontend}, AgentMaster responds with a domain-specific full response to a complex user query by providing a combination of relevant specific information from the database and general information. Figure \ref{fig:demo_backend} displays the coordinator agent decomposing the complex query into sub-questions before assigning each sub-question to the appropriate agents. In the example, the general agent and the SQL agent were employed to generate partial responses, which were sent back to the coordinator to integrate them into a cohesive final response. Additionally, in Figure \ref{fig:demo_check}, the corresponding sub-questions were submitted to AgentMaster to validate the complex query results, and the simple query results were found to be consistent with the information in the complex query responses. AgentMaster was queried for the total number of bridges built in Virginia and those built in Virginia in 2019, and correct information was provided, indicating accurate routing of the complex query and successful SQL database retrieval. Similarly, Figure \ref{fig:complex_query_verfication} in Appendix A.1 displays complex query evaluation, verifying the reliability of AgentMaster. 

\begin{table}[t]
\centering
\small
\renewcommand{\arraystretch}{1.3}
\hfill 
\begin{tabular}{|p{0.6cm}|c|p{3cm}|}
\hline
\textbf{ID} & \textbf{Num of Sub-Questions} & \textbf{Assigned Agents}  \\
\hline 
Q1 & 3 & General, SQL, SQL \\
Q2 & 3 & IR, SQL, SQL \\
Q3 & 5 & IR, SQL, IR, SQL \\
Q4 & 3 & SQL, SQL, IR \\
Q5 & 2 & SQL, General \\
Q6 & 8 & 8 IRs \\
\hline
\end{tabular}

\caption{The number of query decompositions and the corresponding path for each complex query.}
\label{table: c_query_table}
\end{table}

\begin{table}[h]
\centering
\small
\caption{Evaluation Metrics by Query Type}
\label{table:eval_results}
\begin{tabular}{|p{2.725cm}|p{1cm}|p{1.30cm}|p{1cm}|}
\hline
\textbf{Query Type} & \textbf{G-Eval (  \%)} & \textbf{BERT Precision (\%)} & \textbf{BERT F1 (\%)} \\
\hline
SQL Queries & 92.0 & 98.8 & 98.7 \\
\hline
IR Queries & 90.2 & 97.6 & 97.8 \\
\hline
General QA & 84.0 & 95.7 & 96.8 \\
\hline
Image/Complex QA & 82.0 & 90.1 & 91.9 \\
\hline
\textbf{Average} & \textbf{87.1} & \textbf{95.6} & \textbf{96.3} \\
\hline
\end{tabular}
\end{table}

\begin{table*}[h]
\centering
\small
\caption{Feature-Level Comparisons between LLM-Based MAS without A2A or MCP (MAS-0), A2A-Only, MCP-Only, and AgentMaster}
\label{table:comparisons}
\begin{tabular}{|p{3.5cm}|p{2.3cm}|p{2.3cm}|p{2.3cm}|p{2.4cm}|}
\hline
\textbf{Feature} & \textbf{MAS-0} & \textbf{A2A-Only} & \textbf{MCP-Only} & \textbf{AgentMaster    (A2A + MCP)} \\
\hline
Memory & Limited (static or no memory) & Flexible (local and dynamic) & None & Flexible (local and shared) \\
\hline
Coordination & Static & Dynamic & Static & Dynamic \\
\hline
Scalability & Moderate & High & Moderate & High \\
\hline
Fallback Strategy & Limited & Adaptive & Limited & Robust, adaptive \\
\hline
Failure Tolerance & Low & High & Low & High \\
\hline
Inter-Agent Communication & Limited or none & Strong & Limited or none & Strong \\ 
\hline
Architecture & Centralized & Peer-to-peer & Centralized & Hybrid\\
\hline
Task Allocation & Manual or implicit & Decentralized & Centralized & Hybrid\\
\hline
Learning Capability & None or minimal &  Local/online & Centralized/offline & Hybrid: adaptive, distributed, and online \\
\hline
\end{tabular}
\end{table*}

As shown in Table \ref{table: c_query_table}, six complex queries were submitted to AgentMaster. The Coordinator agent performed query decomposition into multiple sub-questions, which were automatically assigned to appropriate agents according to their capabilities. The automated complex actions of the backend -- query decomposition, task allocation, dynamic routing -- as well as resulting outputs are presented in Tables \ref{table: table_backend1}-\ref{table: table_backend2} and Figures \ref{fig:complex_Q1-Q2}-\ref{fig:complex_Q5-Q6} in Appendix A.2. Human evaluation, based on the agentic metrics comprised of task completion and correction, revealed that each complex query was correctly decomposed, with most agent task paths correctly assigned.

\subsection{Overall Evaluation}

The overall A2A-MCP framework was evaluated across multiple dimensions, including factual correctness, relevance, completeness, and semantic similarity. Metrics included Answer Relevancy, Hallucination detection, G-Eval (LLM-based assessment) \citep{liu2023g}, and BERTScore \citep{zhang2024comprehensive}. The test set comprised diverse queries spanning SQL retrieval, IR, general knowledge, and summarization.

Table~\ref{table:eval_results} reports the aggregated metrics across all query types for 23 questions, including both simple and complex questions. Overall, the system demonstrates strong correctness and semantic alignment, with Answer Relevancy and Hallucination metrics indicating high reliability across domains. The average G-Eval score for complex queries exceeds 87.1\%, while BERTScore F1 averages 96.3\%, reflecting high semantic fidelity to reference outputs. 

In the individual agent evaluation, the SQL agent and IR agent produce consistently accurate results, while the general agent and image agent show minor variability due to open-ended generation. Evaluation of complex queries confirms effective decomposition and integration by the Coordinator agent, with most sub-questions yielding outputs consistent with the composite responses.

Table \ref{table:comparisons} presents the summarized feature-based and overall comparisons between existing systems and AgentMaster, respectively, showcasing the comprehensive features of AgentMaster with A2A and MCP.

\section{Conclusions}

This paper presents AgentMaster, a novel modular conversational framework leveraging A2A-MCP protocols for retrieval-augmented question answering across structured, unstructured, and multimodal data sources, facilitating structural clarity and code efficiency as well as easier scalability and maintenance. By interacting with AgentMaster using natural language communication, users can receive domain-specific information regardless of expertise. The experimental results BERTScore F1 and LLM-as-a-Judge metric G-Eval average 96.3\% and 87.1\%, yielding high performance. Validation through both human evaluation and quantitative metrics demonstrates the ability to effectively coordinate various agents, perform complex actions, and produce accurate, semantically faithful responses. The proposed architecture highlights the potential of agent-based orchestration for scalable, domain-adaptive conversational AI.

\section{Limitations}

While the framework achieved strong performance across diverse query types, some limitations remain. The accuracy of retrieval and generation is partly constrained by the underlying LLM and retrieval corpus. Occasional misclassification of query complexity can lead to unnecessary decomposition or incomplete responses. Limited inter-agent collaboration and the constrained size of the database occasionally led to responses with minimal informational depth. The LLM-based reasoning process may encounter challenges in synthesizing complex information. While LLM-as-a-judge evaluation offers scalability and efficiency, it remains limited by potential biases, lack of task-specific expertise, and alignment with human judgment. Finally, the current framework lacks established security safeguards for information storage and usage. These limitations can be addressed in future work.

\section{Acknowledgments}

The authors thank the Federal Highway Administration (FHWA) for providing public datasets used to build knowledge databases for the case study. The authors also thank Ellie L. Zhang for contributing to framework building.

\bibliography{custom}

\onecolumn
\appendix

\section{Appendix}

The appendix contains more demonstrations of multimodal inputs and outputs, single and complex queries, verification, automated actions (query decomposition, task allocation, and dynamic routing) for complex queries in the backend, as well as different usages of the domain-specific or specialized agents.

\subsection{AgentMaster Query Demonstrations}

\begin{figure*}[h]
    \centering
    \includegraphics[height=3.3in]{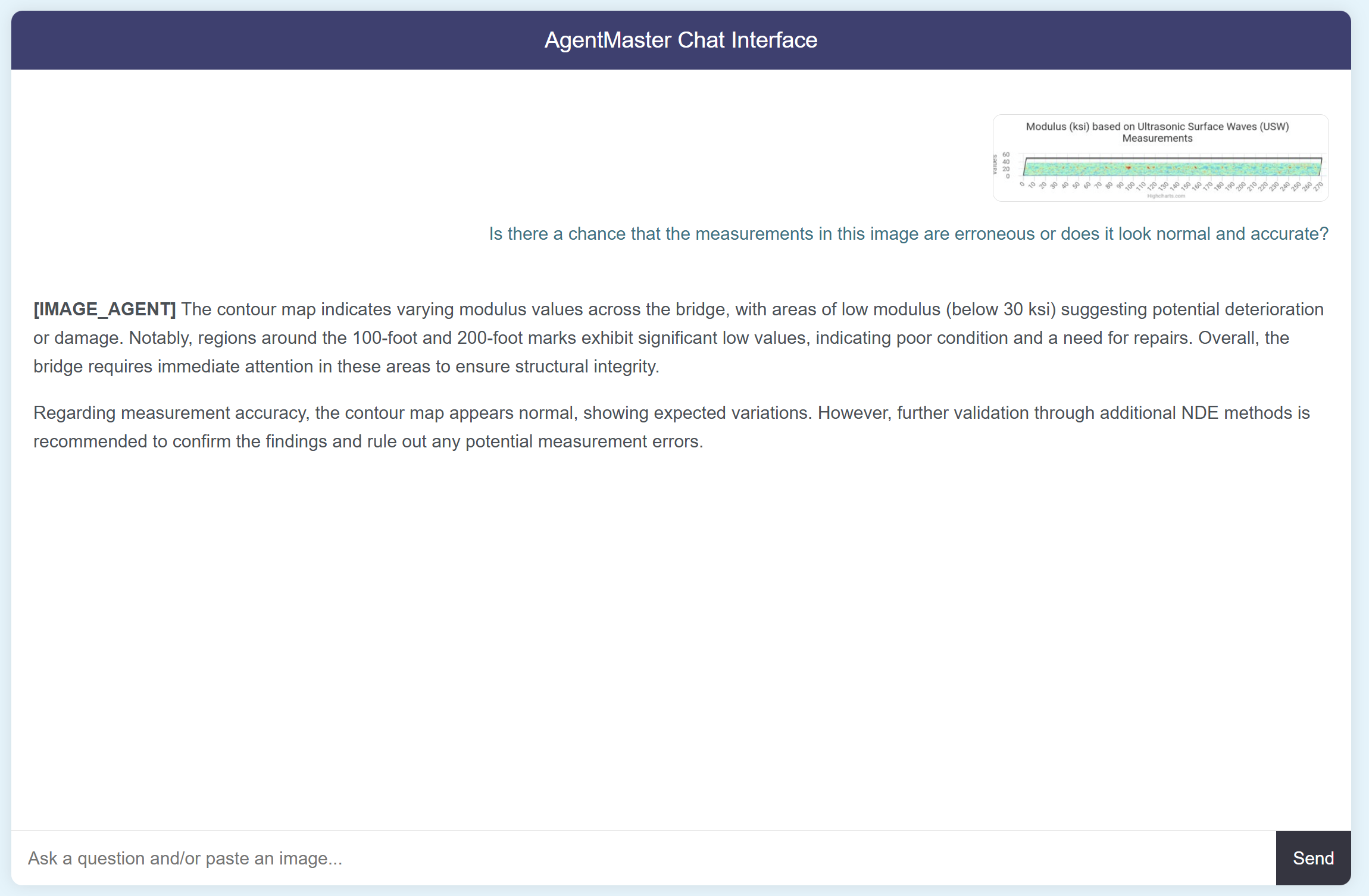}
    \caption{Frontend example of a single query for the Image agent.}
    \label{fig:image_frontend_ex}
\end{figure*}

\begin{figure}[h]
    \centering
    \includegraphics[height=3.3in]{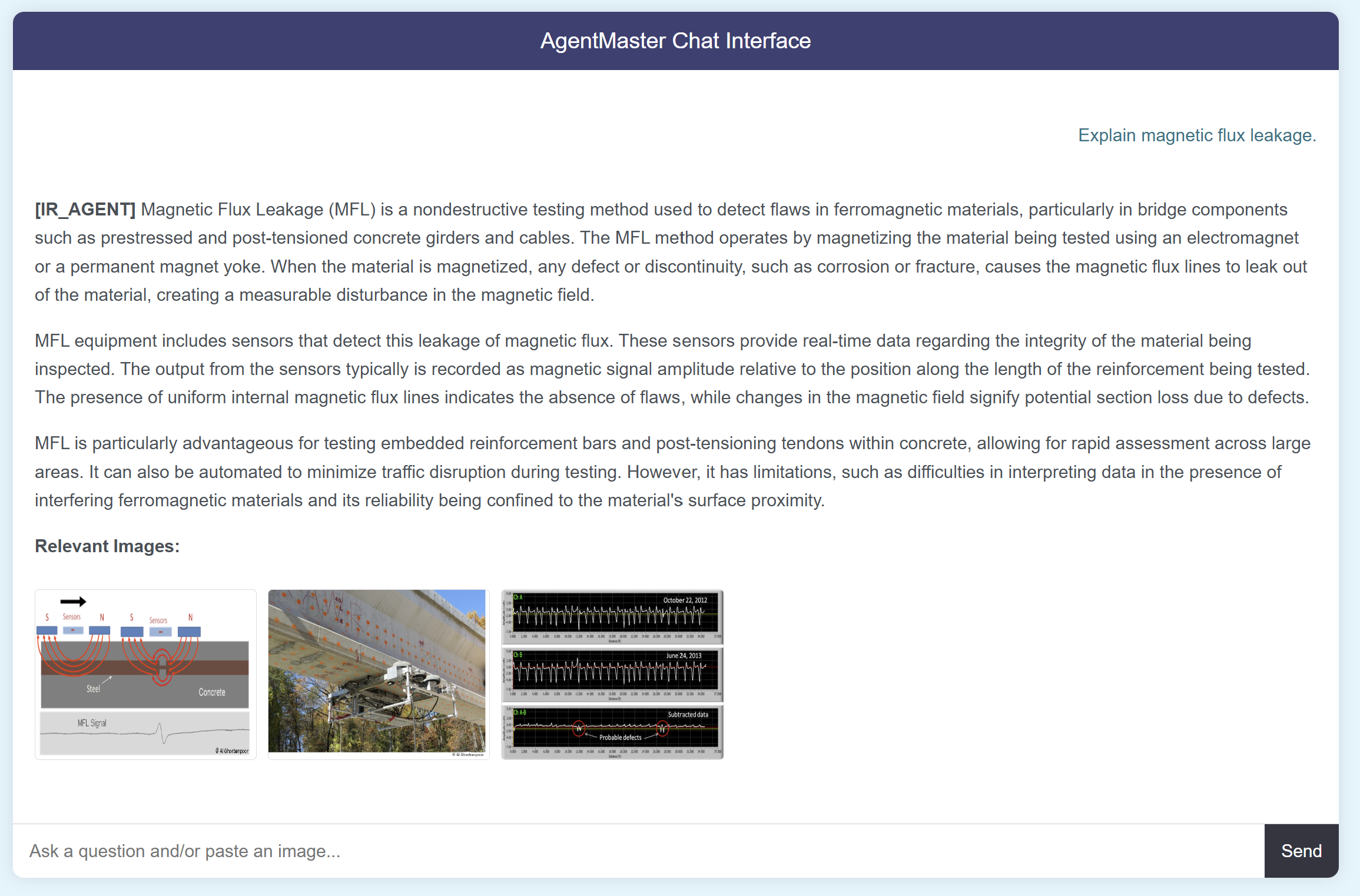}
    \caption{Frontend example of a single query for the IR agent.}
    \label{fig:mag_flux_frontend_ex}
\end{figure}

\begin{figure*}[t!]
    \centering
    \begin{subfigure}[t]{\textwidth}
        \centering
        \includegraphics[width=0.85\columnwidth]{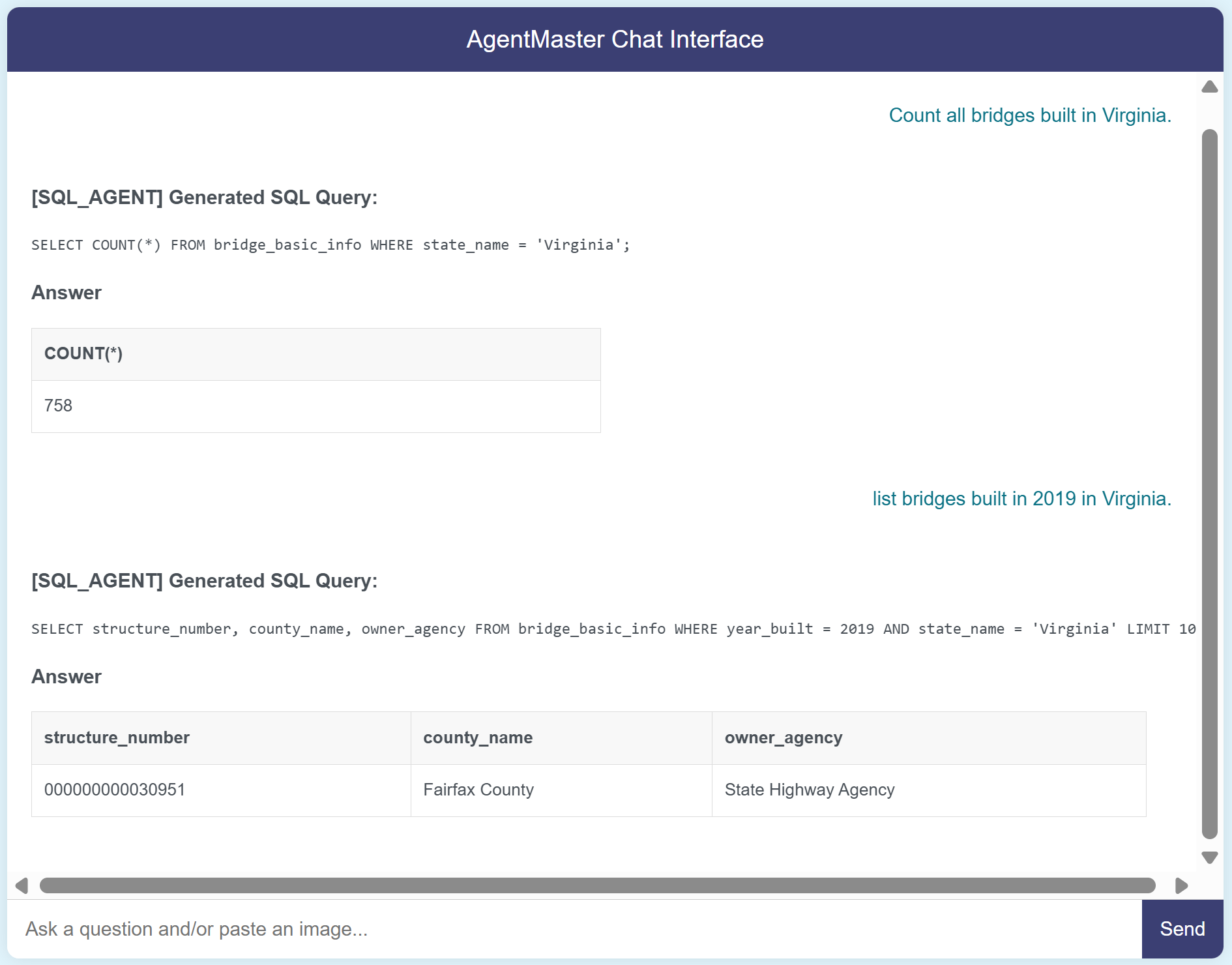}
        \caption{Frontend example of two single queries for the SQL agent.}
        \label{fig:demo_queries_traffic1}
    \end{subfigure}%
    
    \centering
    \begin{subfigure}[t]{\textwidth}
        \centering
        \includegraphics[width=0.85\columnwidth]{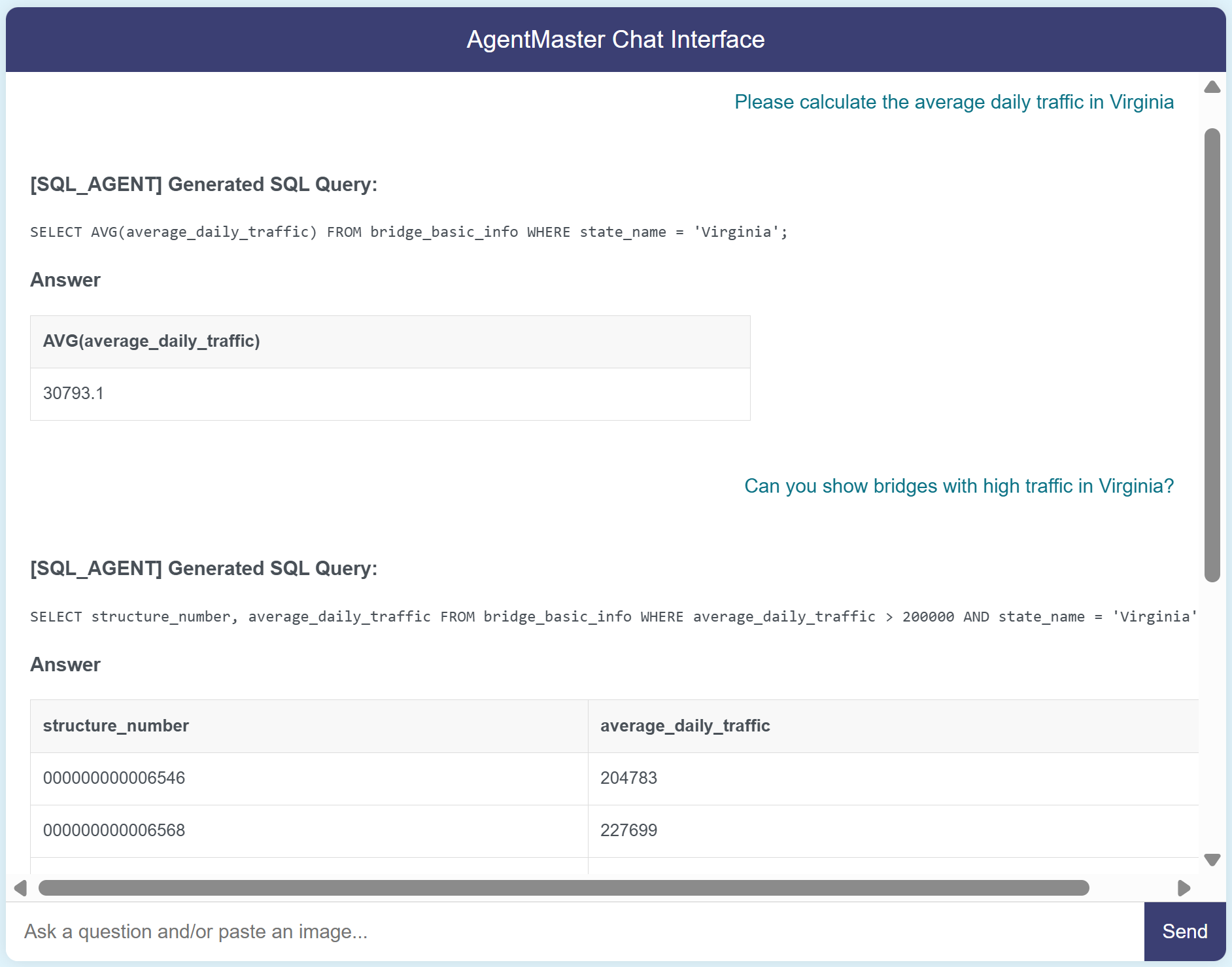}
        \caption{Frontend example of two single queries for the SQL agent.}
        \label{fig:demo_queries_traffic2}
    \end{subfigure}%

    \caption{More demonstration examples of a single query for the SQL agent.}
    \label{fig:more_demos_SQL1}
\end{figure*}

\begin{figure*}[t!]
    \centering    
    \begin{subfigure}[t]{\textwidth}
        \centering
        \includegraphics[width=0.9\columnwidth]{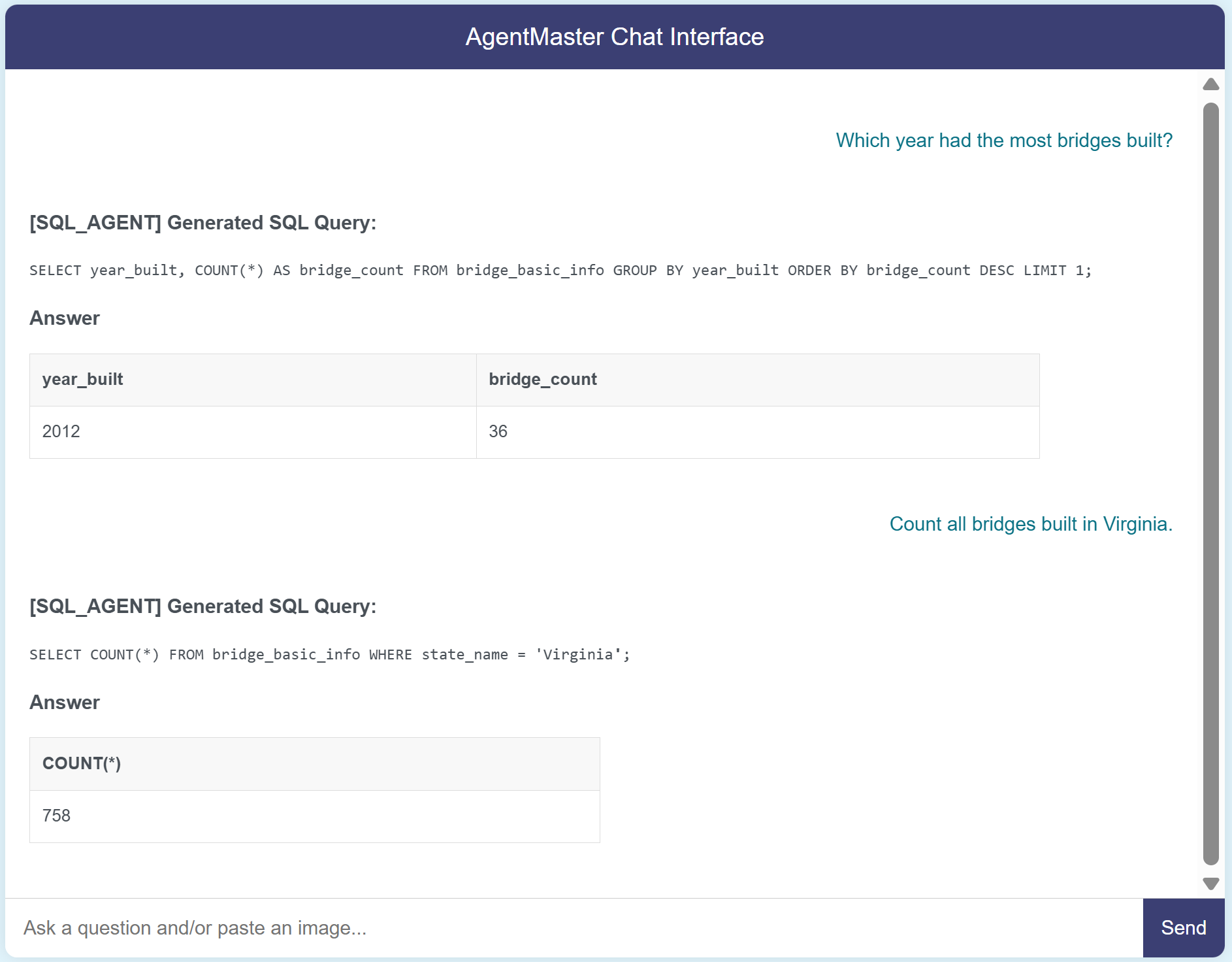}
        \caption{Frontend example of two single queries for the SQL agent.}
        \label{fig:demo_queries_SQL1}
    \end{subfigure}%
    
    \begin{subfigure}[t]{\textwidth}
        \centering
        \includegraphics[width=0.9\columnwidth]{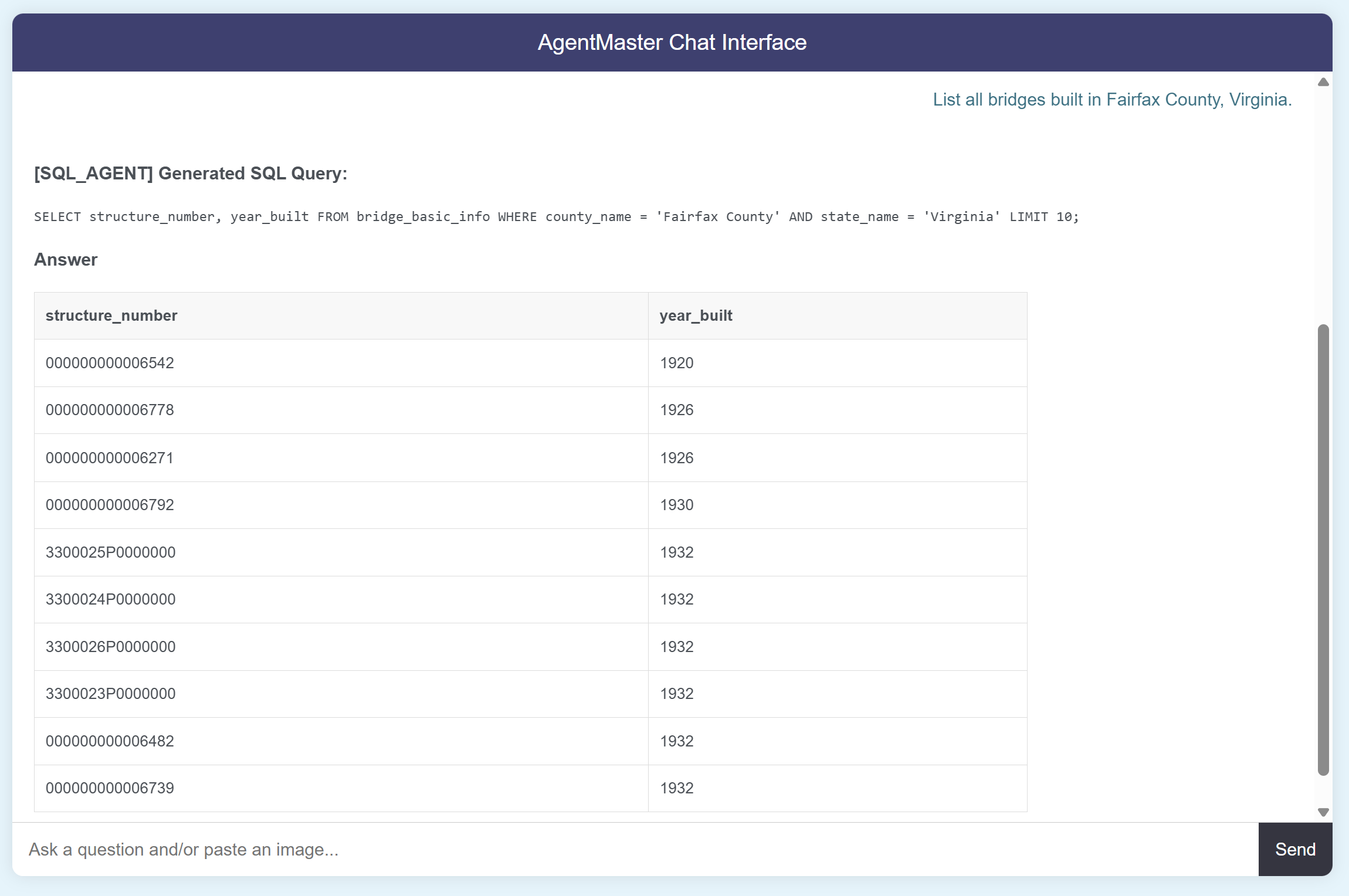}
        \caption{Frontend example of one single query for the SQL agent.}
        \label{fig:demo_queries_SQL2}
    \end{subfigure}

    \caption{More demonstration examples of a single query for the SQL agent.}
    \label{fig:more_demos_SQL2}
\end{figure*}

\begin{figure*}[t!]
    \begin{subfigure}[t]{\textwidth}
        \centering
        \includegraphics[width=0.9\columnwidth]{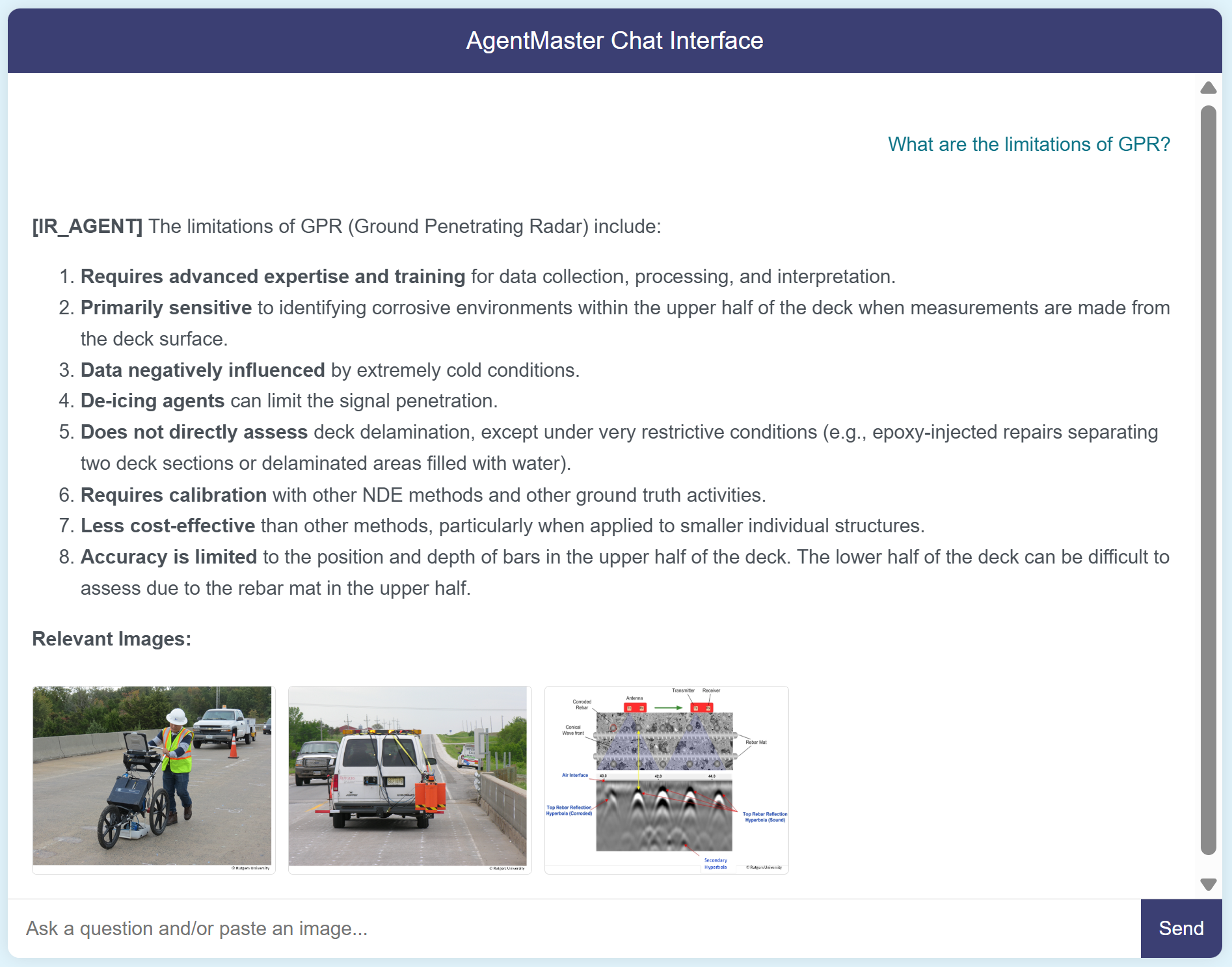}
        \caption{A single query for the IR agent.}
        \label{fig:demo_traffic_verify1}
    \end{subfigure}%
    
    \begin{subfigure}[t]{\textwidth}
        \centering
        \includegraphics[width=0.9\columnwidth]{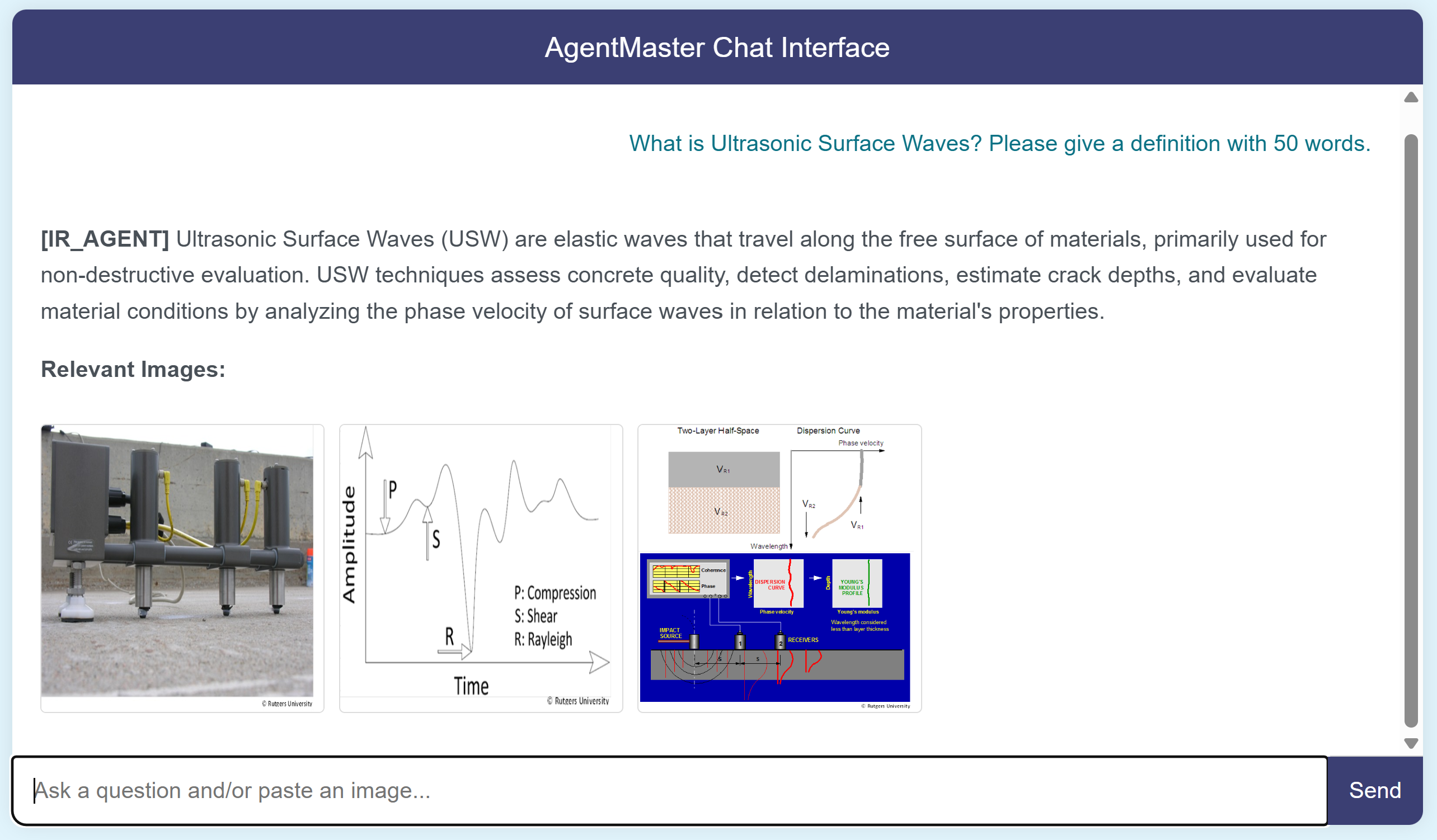}
        \caption{A single query of the IR agent.}
        \label{fig:demo_traffic_verify2}
    \end{subfigure}
    \caption{Two examples of single verification queries for the IR agent.}
    \label{fig:demo_traffic}
\end{figure*}

\begin{figure*}[t!]
    \begin{subfigure}[t]{\textwidth}
        \centering
        \includegraphics[width=0.9\columnwidth]{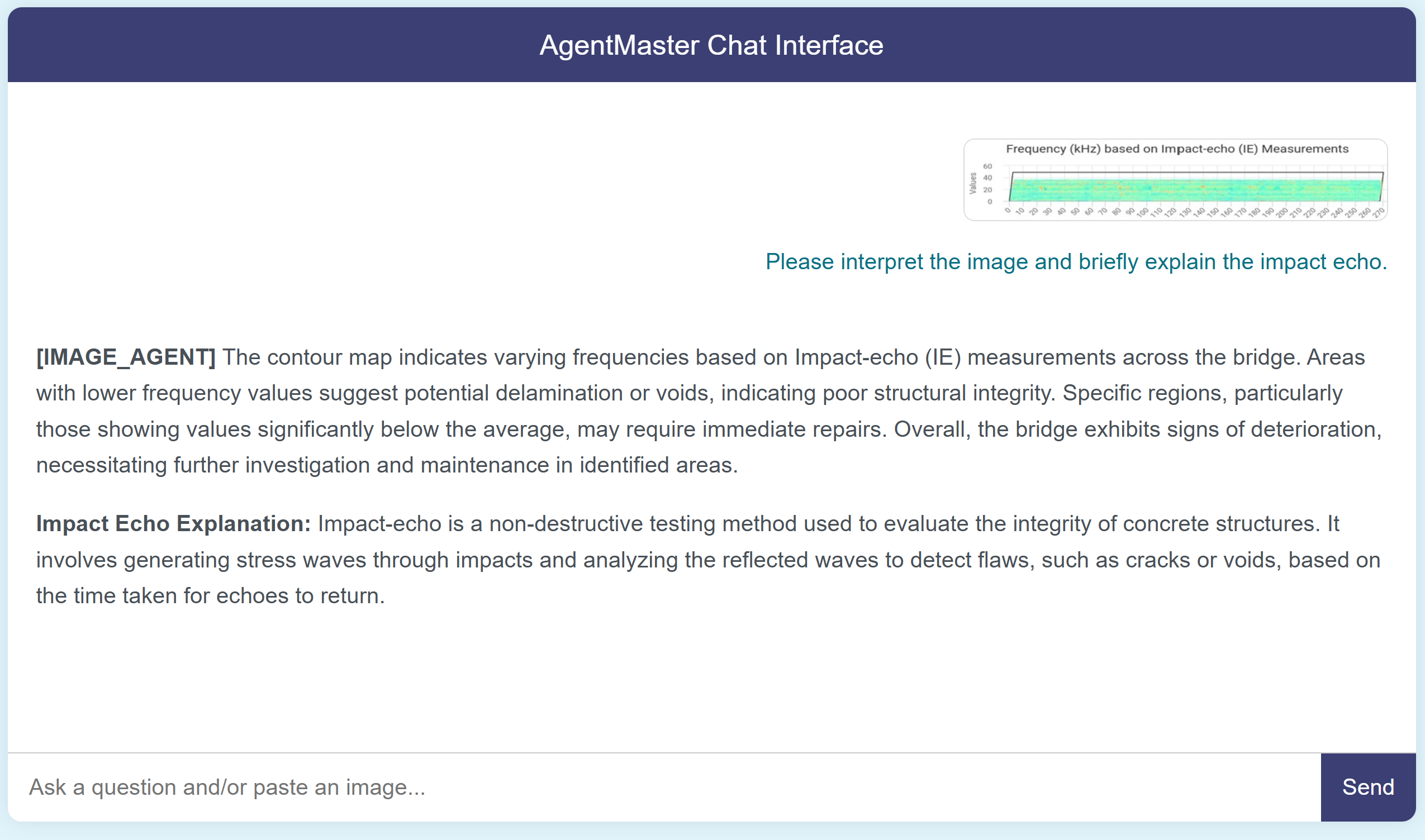}
        \caption{A single query for the image agent.}
        \label{fig:demo_traffic_verify3}
    \end{subfigure}%
    
    \begin{subfigure}[t]{\textwidth}
        \centering
        \includegraphics[width=0.9\columnwidth]{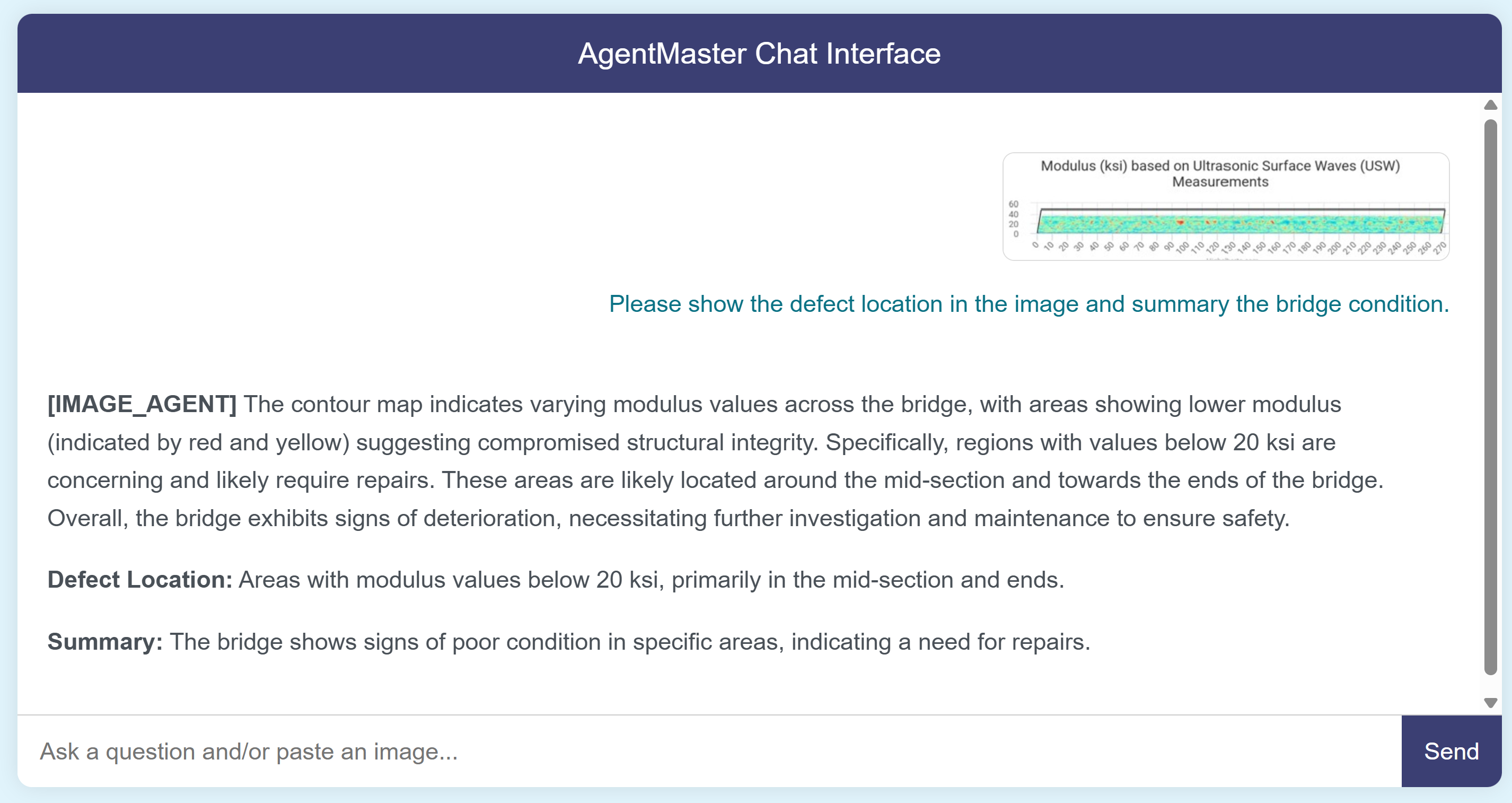}
        \caption{A single query for the image agent.}
        \label{fig:demo_traffic_verify4}
    \end{subfigure}
    \caption{Two examples of single verification queries for the image agent.}
    \label{fig:demo_traffic}
\end{figure*}

\begin{figure*}[t!]
    \centering
     
    \begin{subfigure}[t]{\textwidth}
        \centering
        \includegraphics[width=0.9\columnwidth]{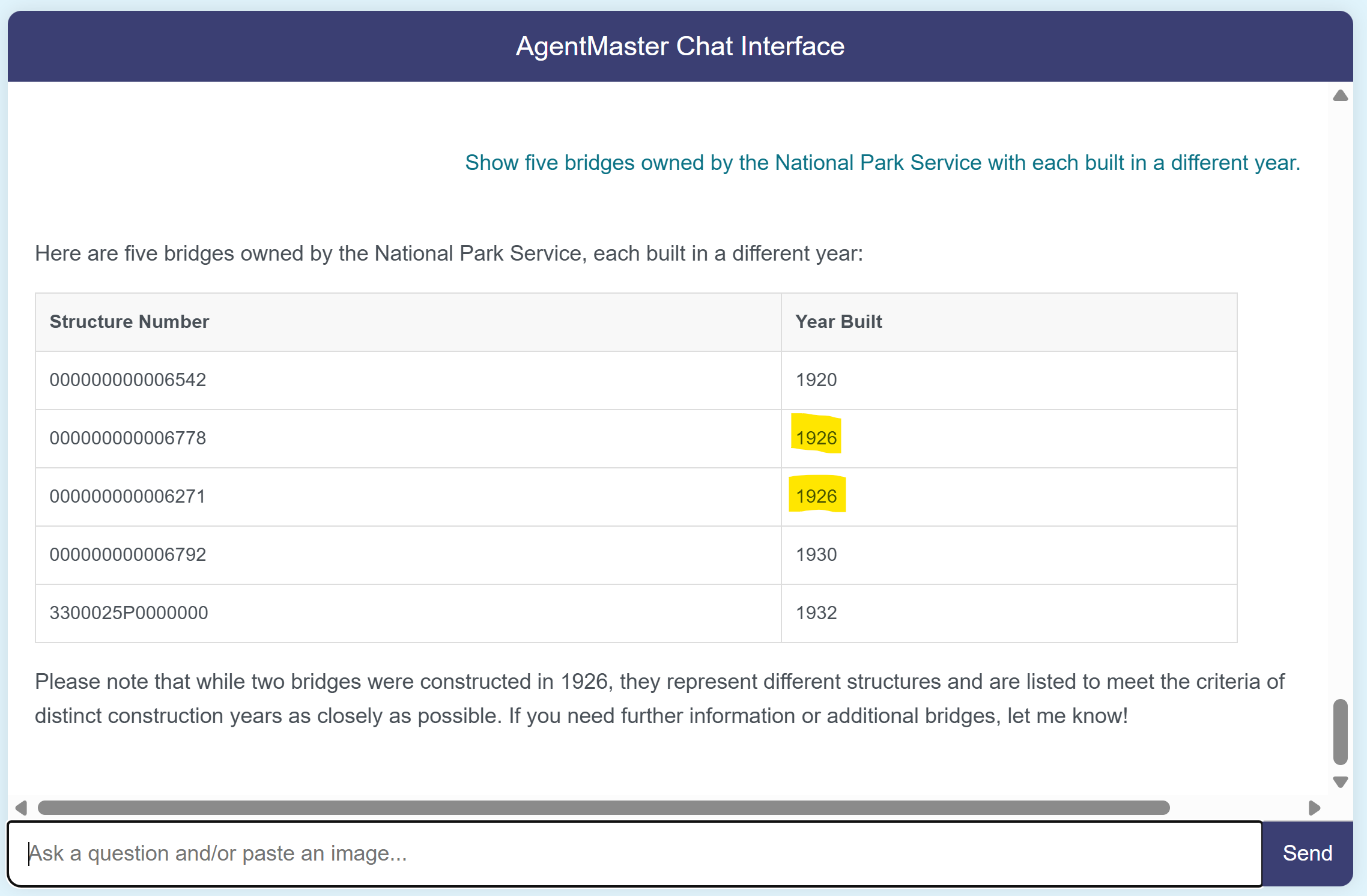}
        \caption{SQL agent error from a single query.}
        \label{fig:demo_traffic_verify5}
    \end{subfigure}%
    
    \begin{subfigure}[t]{\textwidth}
        \centering
        \includegraphics[width=0.9\columnwidth]{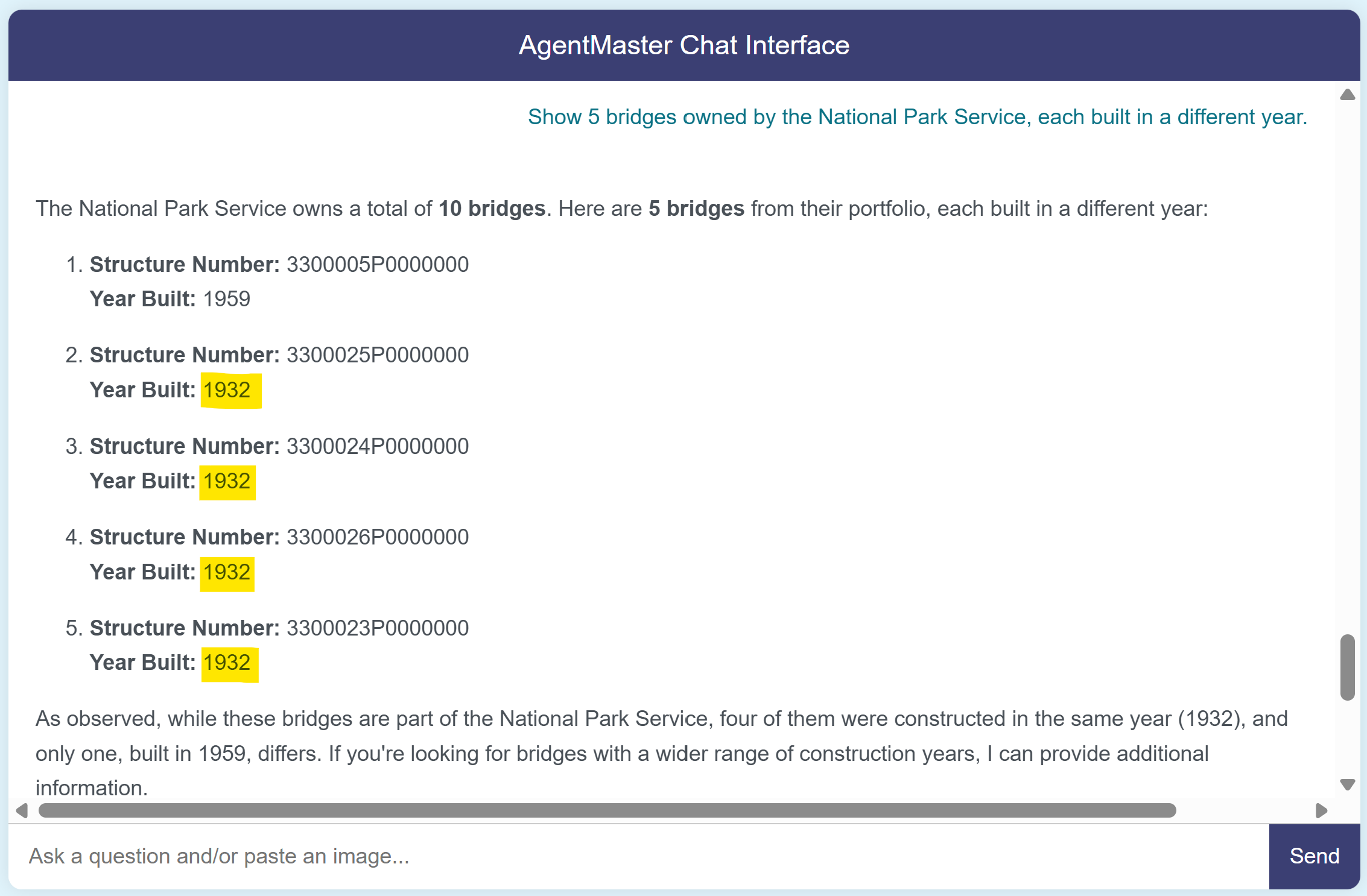}
        \caption{SQL agent error from a single query.}
        \label{fig:demo_traffic_verify6}
    \end{subfigure}
    \caption{The highlighted text contains errors in SQL information retrieval, including duplicate content, indicating that AgentMaster may still produce erroneous results. }
    \label{fig:demo_traffic}
\end{figure*}

\newpage

\begin{figure*}[t!]
    \centering
    \begin{subfigure}[t]{\textwidth}
        \centering
        \includegraphics[height=4.4in]{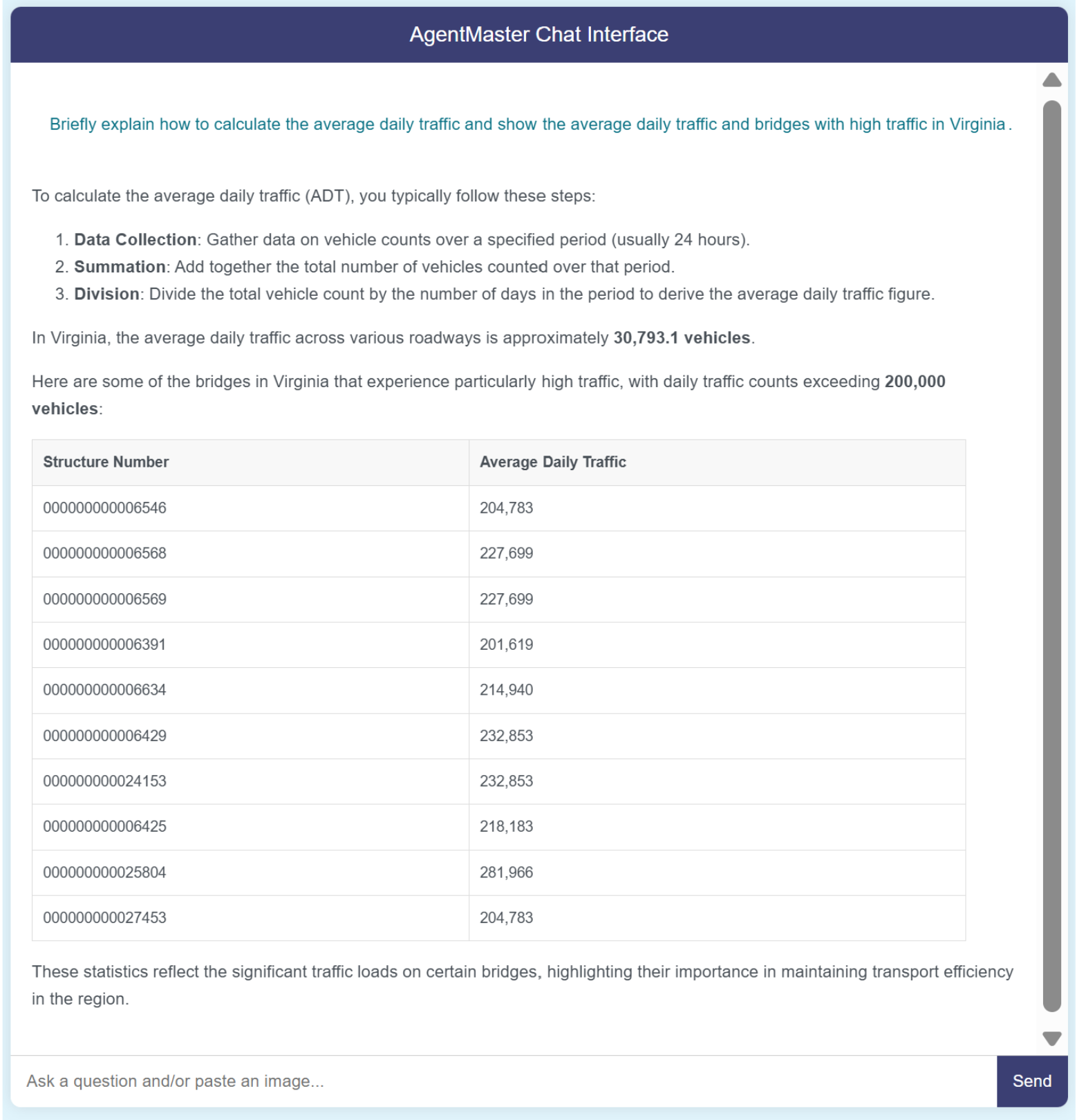}
        \caption{Example of a complex query.}
        \label{fig:demo_traffic_query}
    \end{subfigure}%
     
    \begin{subfigure}[t]{\textwidth}
        \centering
        \includegraphics[height=4.4in]{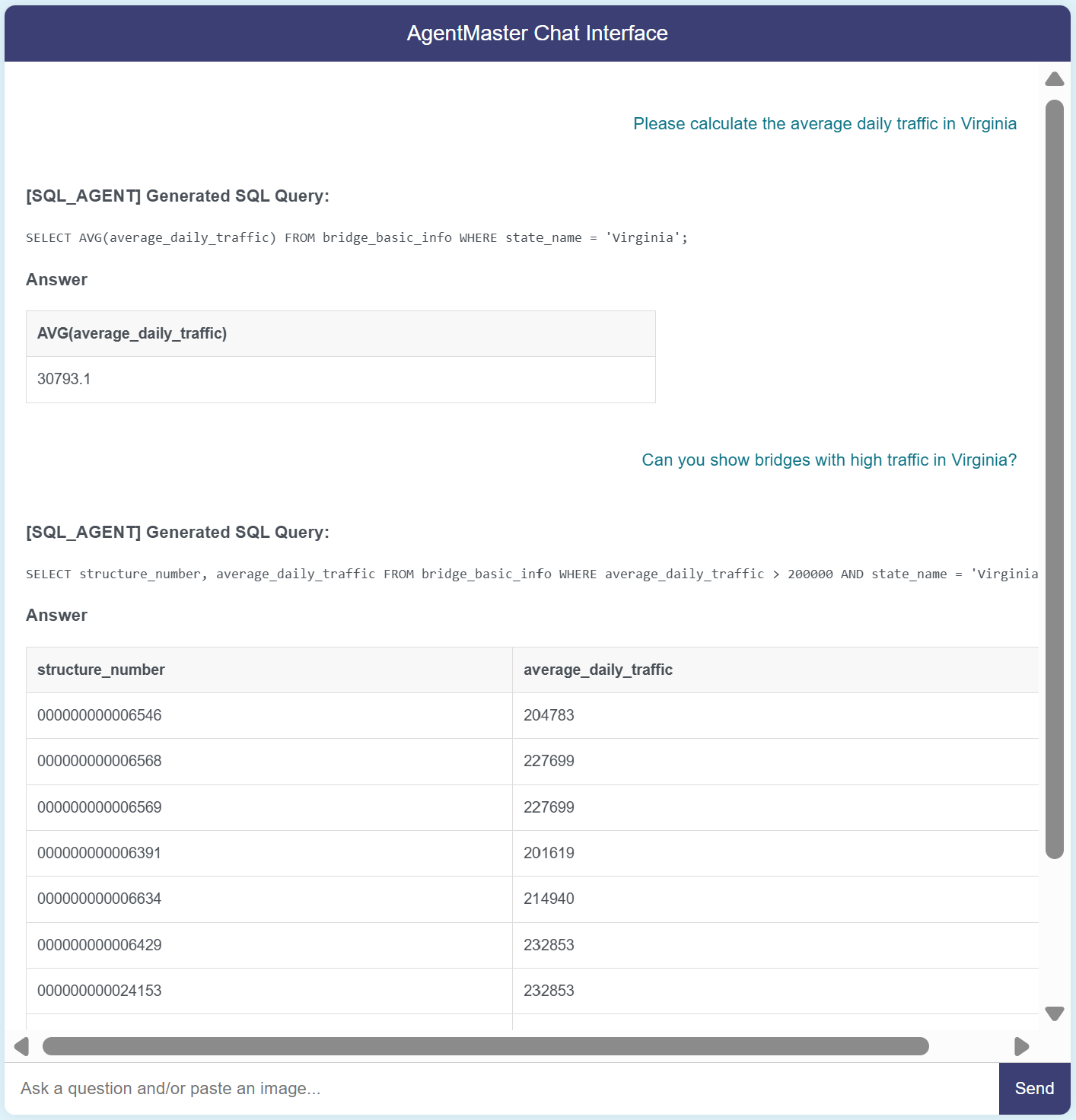}
        \caption{Single queries for verification.}
        \label{fig:demo_traffic_verify7}
    \end{subfigure}
    \caption{An additional AgentMaster demonstration example and verification.}
    \label{fig:complex_query_verfication}
\end{figure*}

\clearpage
\subsection{Decomposition, Allocation, and Routing of Agent Tasks for Complex Queries}

\subsubsection{Complex Queries}
\begin{table*}[h!]
\centering
\small
\renewcommand{\arraystretch}{1.3}
\hfill 
\begin{tabular}{| c | p{15cm}|}
\hline
\textbf{ID} & \textbf{Complex Query}  \\
\hline 
Q1 & Briefly define a bridge, then provide the total number of bridges in Virginia and list those built in 2019. \\
Q2 & Briefly explain how to calculate the average daily traffic and show the average daily traffic and bridges with high traffic in Virginia. \\
Q3 & What is the typical service lifespan of a bridge? Identify and display all bridges that exceed this average age. \\
Q4 & List the three oldest bridges in Virginia, show their year built from the database, and briefly explain why their maintenance costs tend to be higher according to engineering guidelines with 50 words. \\
Q5 & List five of the oldest bridges in the United States still in use today, and briefly describe their historical significance. \\
Q6 & Compare the advantages and disadvantages of concrete arch bridges and steel truss bridges in terms of maintenance, lifespan, and load capacity. \\
\hline
\end{tabular}

\caption{The complex questions and the corresponding IDs, matching Table \ref{table: c_query_table}.}
\label{table: table_questions}
\end{table*}

\subsubsection{AgentMaster Outputs of Complex Queries in the Chat Interface}

\begin{figure*}[h]
    \centering
    \begin{subfigure}[t]{\textwidth}
        \centering
       \includegraphics[width=0.6\columnwidth]{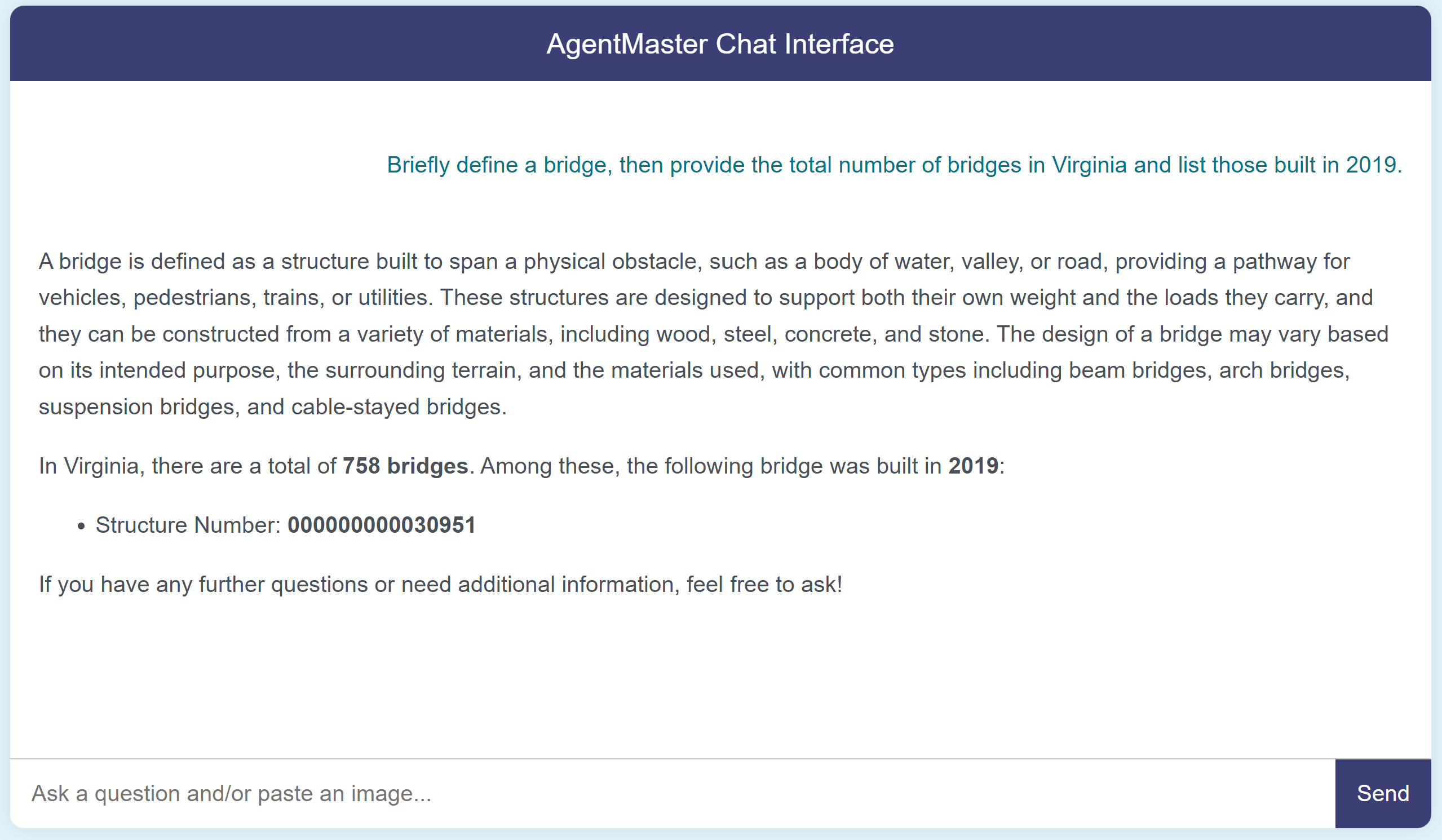}
        \caption{Complex query Q1.}
        \label{fig:demo_traffic_query}
    \end{subfigure}%
     
    \begin{subfigure}[t]{\textwidth}
        \centering
       \includegraphics[width=0.6\columnwidth]{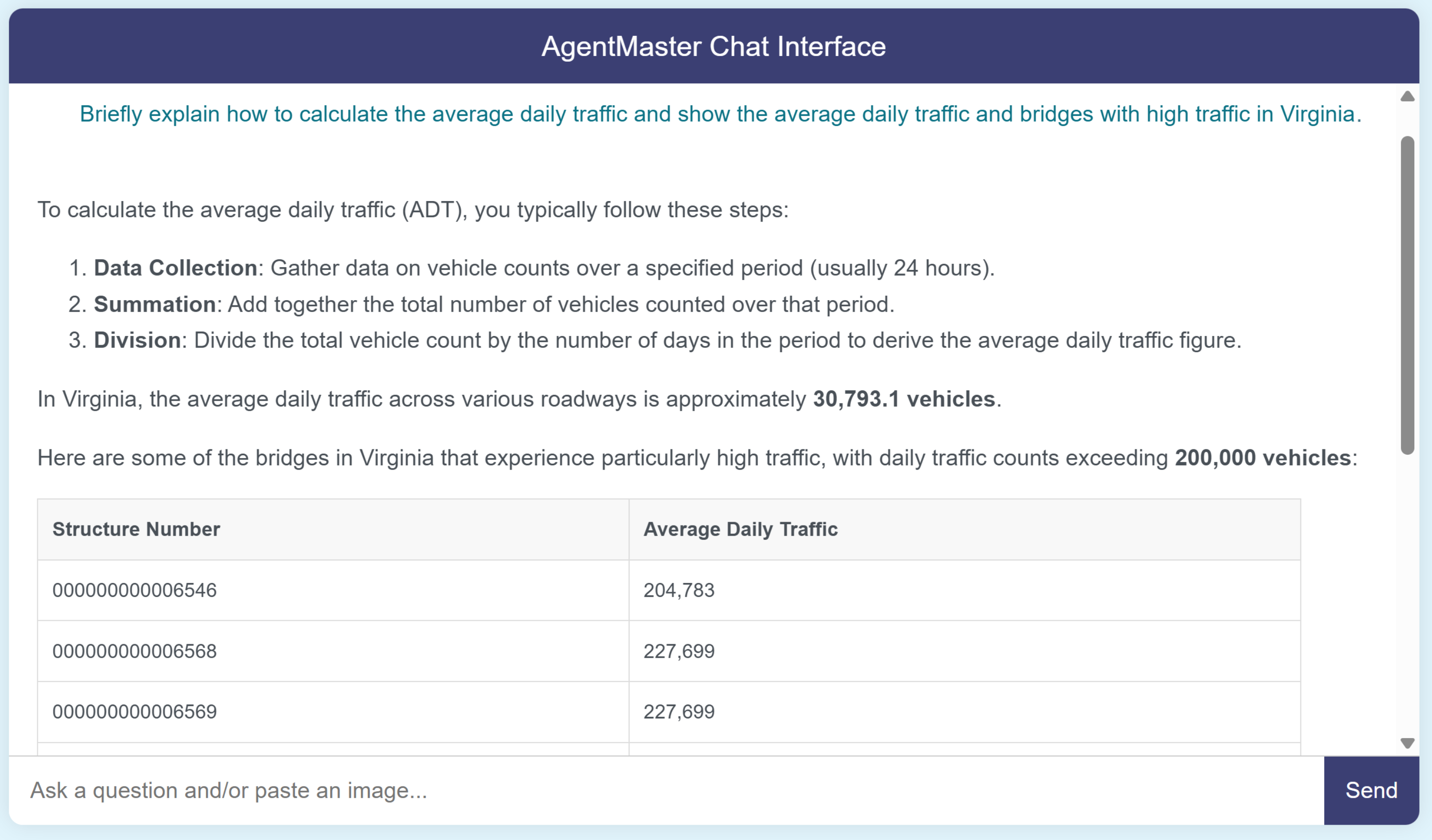}
        \caption{Complex query Q2.}
        \label{fig:demo_traffic_verify8}
    \end{subfigure}
    \caption{The display of the frontend response for complex queries Q1 and Q2.}
    \label{fig:complex_Q1-Q2}
\end{figure*}

\begin{figure*}[h]
    \centering
    \begin{subfigure}[t]{\textwidth}
        \centering
        \includegraphics[width=0.7\columnwidth]
       {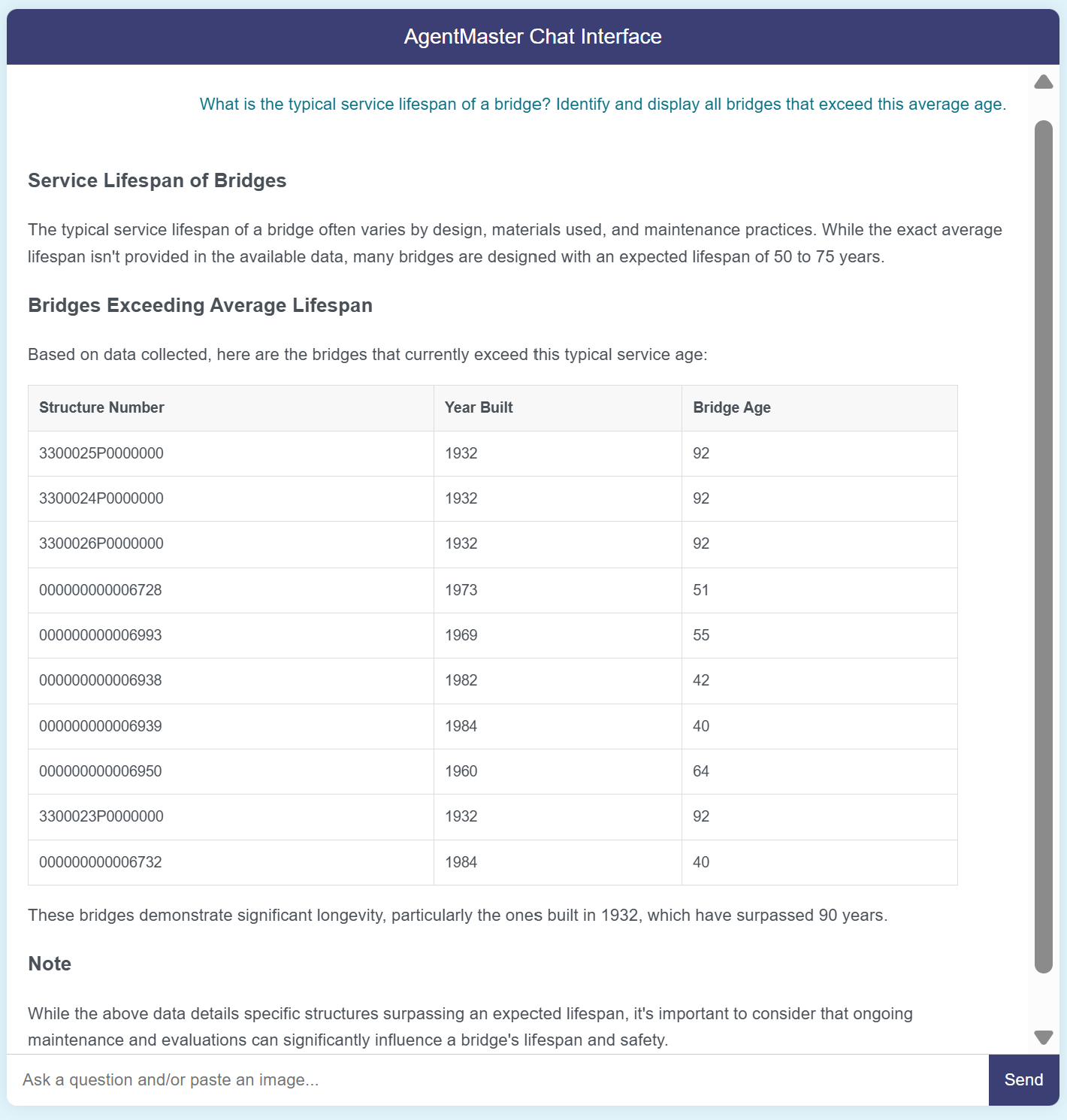}
        \caption{Complex query Q3.}
        \label{fig:demo_traffic_query}
    \end{subfigure}%
    
    \begin{subfigure}[t]{\textwidth}
        \centering
        \includegraphics[width=0.7\columnwidth]
        {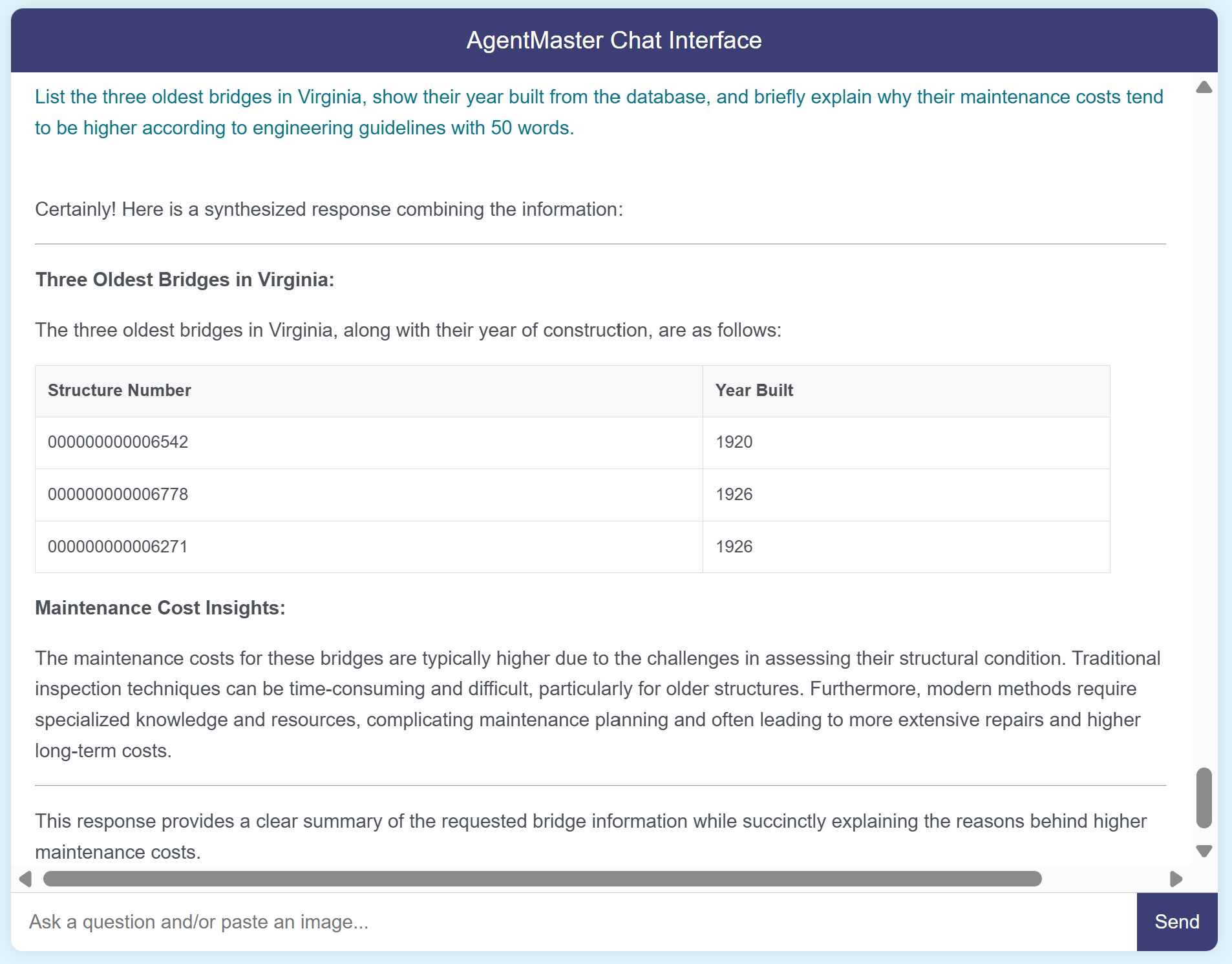}
        \caption{Complex query Q4.}
        \label{fig:demo_traffic_verify9}
    \end{subfigure}
    \caption{The display of the frontend response for complex queries Q3 and Q4.}
    \label{fig:complex_Q3-Q4}
\end{figure*}

\begin{figure*}[h]
    \centering
    \begin{subfigure}[t]{0.6\textwidth}
        \centering
        \includegraphics[width=\columnwidth]
      {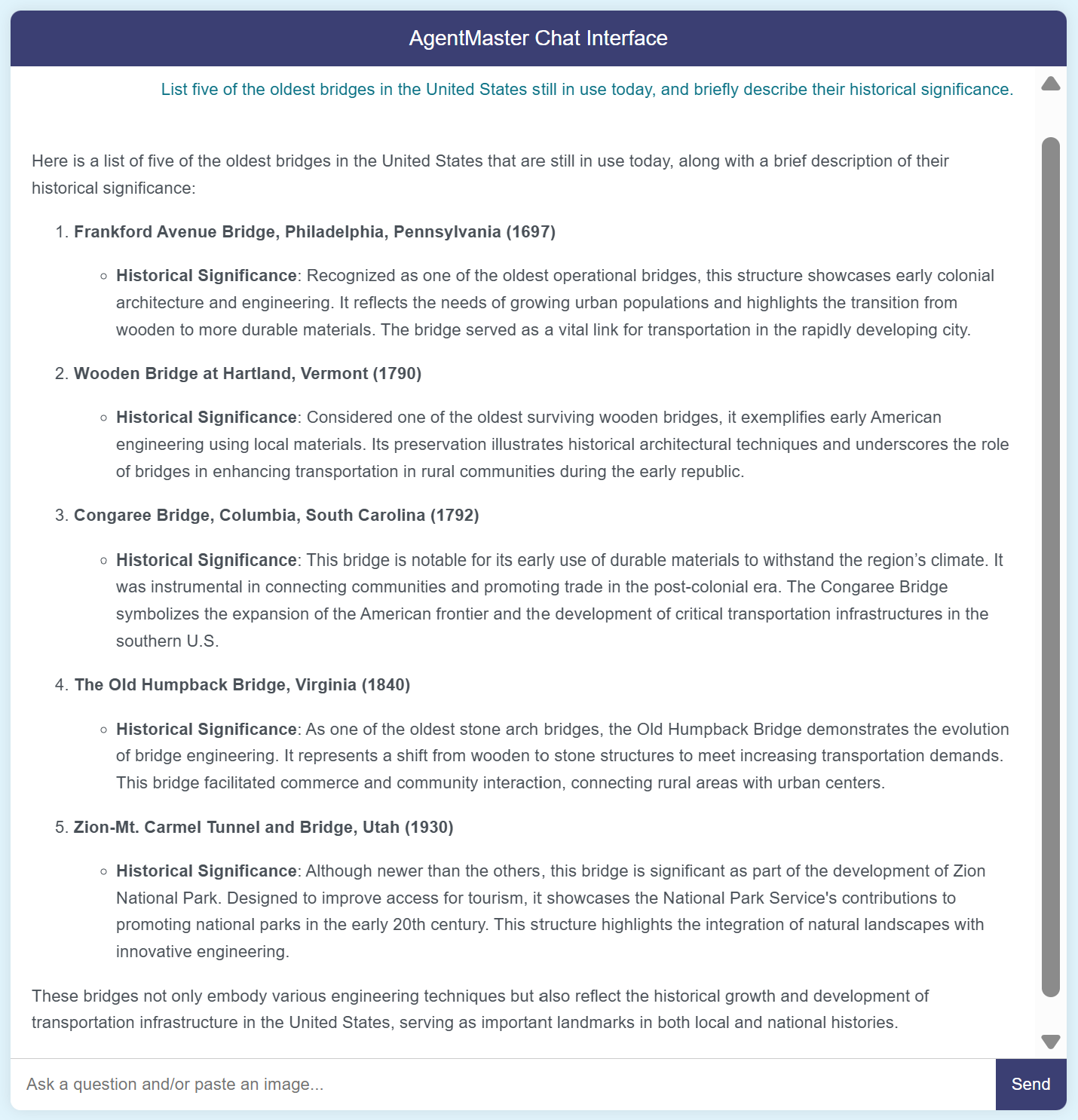}
        \caption{Complex query Q5.}
        \label{fig:demo_traffic_query}
    \end{subfigure}%
    
    \begin{subfigure}[t]{0.6\textwidth}
        \centering
       \includegraphics[width=\columnwidth]
    {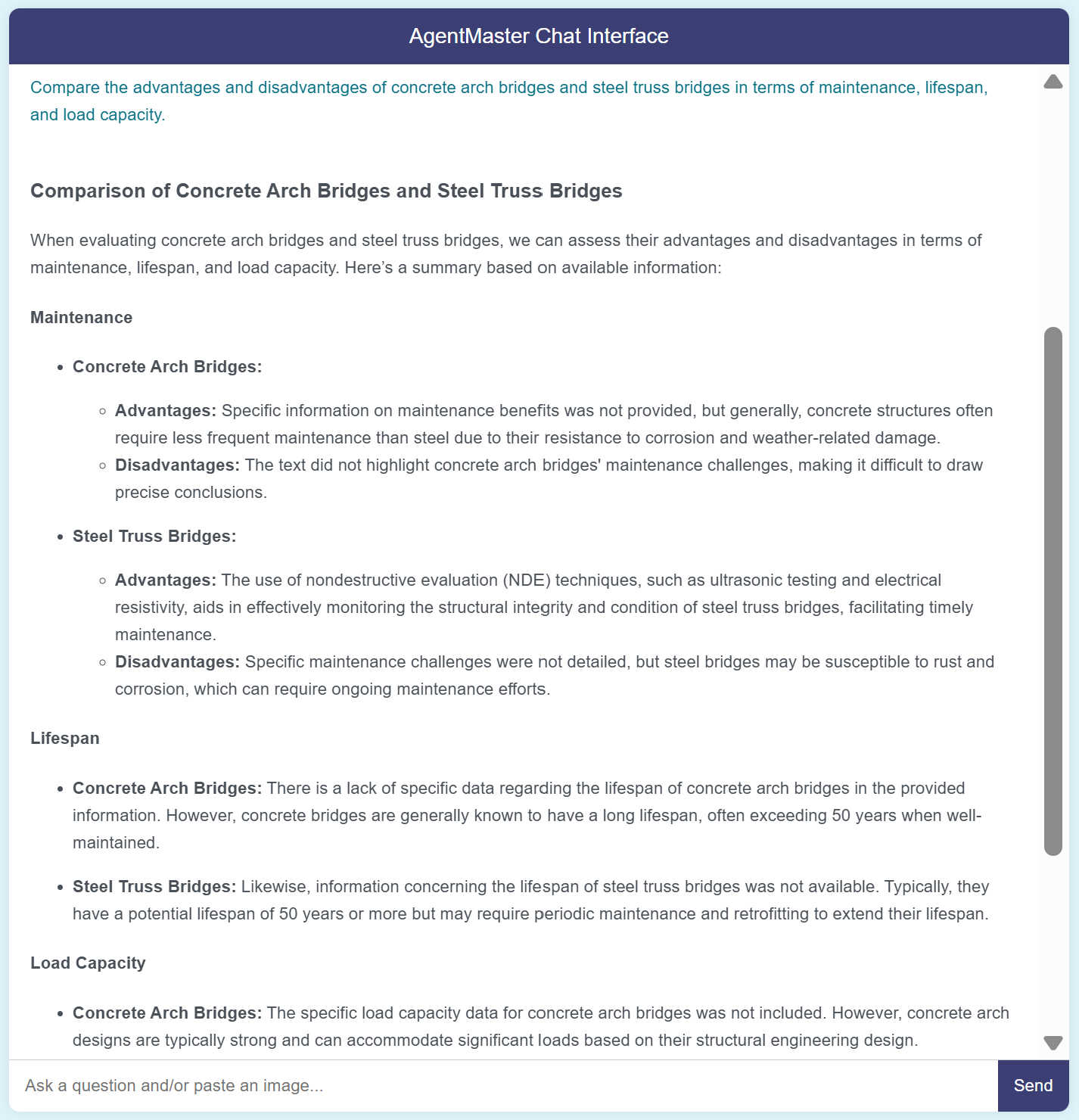}
        \caption{Complex query Q6.}
        \label{fig:demo_traffic_verify10}
    \end{subfigure}
    \caption{The display of the frontend response for complex queries questions Q5 and Q6.}
    \label{fig:complex_Q5-Q6}
\end{figure*}

\clearpage

\subsubsection{Automated Actions (Query Decomposition, Task Allocation, and Dynamic Routing) for Complex Queries in the Backend}

\begin{table*}[h!]
\centering
\small
\renewcommand{\arraystretch}{1.3}
\hfill 
\begin{tabular}{| c | p{14cm}|}
\hline
\textbf{ID} & \textbf{Information for Query Decomposition, Task Allocation, and Dynamic Routing} \\
\hline 
Q1 & [Agent Server] Received complex query for coordinator: `Briefly define a bridge, then provide the total number of bridges in Virginia and list those built in 2019.'
\newline [Coordinator] Decomposing user query...
\newline [Coordinator] Decomposed into: [`What is a brief definition of a bridge?', `What is the total number of bridges in Virginia?', `Which bridges in Virginia were built in 2019?']
\newline [Coordinator] Routing sub-question: `What is a brief definition of a bridge?'
\newline [Complex to Single Query Router] Decision for GENERAL\_AGENT
\newline [Coordinator] Routing sub-question: `What is the total number of bridges in Virginia?'
\newline [Complex to Single Query Router] Decision for SQL\_AGENT
Raw LLM output: "SELECT COUNT(*) FROM bridge\_basic\_info WHERE state\_name = `Virginia';"
\newline Detected valid SQL, executing...
\newline [Coordinator] Routing sub-question: `Which bridges in Virginia were built in 2019?'
\newline [Complex to Single Query Router] Decision for SQL\_AGENT
\newline Raw LLM output: "SELECT structure\_number, year\_built FROM bridge\_basic\_info WHERE state\_name = `Virginia' AND year\_built = 2019 LIMIT 10;"
Detected valid SQL, executing...
\newline [Coordinator] Synthesizing final answer...
 \\
\hline
Q2 & [Agent Server] Received complex query for coordinator: `Briefly explain how to calculate the average daily traffic and show the average daily traffic and bridges with high traffic in Virginia?'
\newline [Coordinator] Decomposing user query...
\newline [Coordinator] Decomposed into: [`How do you calculate the average daily traffic?', `What is the average daily traffic in Virginia?', `Which bridges in Virginia have high traffic?']
\newline [Coordinator] Routing sub-question: `How do you calculate the average daily traffic?'
\newline [Complex to Single Query Router] Decision for IR\_AGENT
\newline [Coordinator] Routing sub-question: `What is the average daily traffic in Virginia?'
\newline [Complex to Single Query Router] Decision for SQL\_AGENT
\newline Raw LLM output: "SELECT AVG(average\_daily\_traffic) FROM bridge\_basic\_info WHERE state\_name = `Virginia';"
\newline Detected valid SQL, executing...
\newline [Coordinator] Routing sub-question: `Which bridges in Virginia have high traffic?'
\newline [Complex to Single Query Router] Decision for SQL\_AGENT
\newline Raw LLM output: "SELECT structure\_number, average\_daily\_traffic FROM bridge\_basic\_info WHERE state\_name = `Virginia' AND average\_daily\_traffic > 200000 LIMIT 10;"
\newline Detected valid SQL, executing...
\newline [Coordinator] Synthesizing final answer...
 \\
\hline
Q3 & [Agent Server] Received complex query for coordinator: `What is the typical service lifespan of a bridge? Identify and display all bridges that exceed this average age.'
\newline [Coordinator] Decomposing user query...
\newline [Coordinator] Decomposed into: [`What is the typical service lifespan of a bridge?', `What is the average age of bridges?', `What bridges exceed the typical service lifespan?', `How can I display the bridges that exceed the average age?']
\newline [Coordinator] Routing sub-question: `What is the typical service lifespan of a bridge?'
\newline [Complex to Single Query Router] Decision for IR\_AGENT
\newline [Coordinator] Routing sub-question: `What is the average age of bridges?'
\newline [Complex to Single Query Router] Decision for SQL\_AGENT
\newline Raw LLM output: `The average age of bridges cannot be determined with a single SQL query as it requires an aggregate function.'
No SQL detected, returning text answer.
\newline [Coordinator] Routing sub-question: `What bridges exceed the typical service lifespan?'
\newline [Complex to Single Query Router] Decision for IR\_AGENT
\newline [Coordinator] Routing sub-question: `How can I display the bridges that exceed the average age?'
\newline [Complex to Single Query Router] Decision for SQL\_AGENT
\newline Raw LLM output: `SELECT structure\_number, year\_built, bridge\_age FROM bridge\_basic\_info WHERE bridge\_age > (SELECT AVG(bridge\_age) FROM bridge\_basic\_info) LIMIT 10;'
\newline Detected valid SQL, executing...
\newline [Coordinator] Synthesizing final answer...
 \\
\hline
\end{tabular}

\caption{Automated actions for complex queries (Q1-Q3) in the backend.}
\label{table: table_backend1}
\end{table*}

\begin{table*}[h!]
\centering
\small
\renewcommand{\arraystretch}{1.3}
\hfill 
\begin{tabular}{| c | p{14cm}|}
\hline
\textbf{ID} & \textbf{Information for Query Decomposition, Task Allocation, and Dynamic Routing} \\
\hline 
Q4 & [Agent Server] Received complex query for coordinator: `List the three oldest bridges in Virginia, show their year built from the database, and briefly explain why their maintenance costs tend to be higher according to engineering guidelines with 50 words.' 
\newline [Coordinator] Decomposing user query...
\newline [Coordinator] Decomposed into: [`What are the three oldest bridges in Virginia?', `What is the year built for each of the three oldest bridges in Virginia?', `Why do the maintenance costs of these bridges tend to be higher according to engineering guidelines?']
\newline[Coordinator] Routing sub-question: `What are the three oldest bridges in Virginia?'
\newline[Complex to Single Query Router] Decision for SQL\_AGENT
\newline Raw LLM output: "SELECT structure\_number, year\_built FROM bridge\_basic\_info WHERE state\_name = `Virginia' ORDER BY year\_built ASC LIMIT 3;"
\newline Detected valid SQL, executing...
\newline[Coordinator] Routing sub-question: `What is the year built for each of the three oldest bridges in Virginia?'
\newline [Complex to Single Query Router] Decision for SQL\_AGENT
\newline Raw LLM output: "SELECT structure\_number, year\_built FROM bridge\_basic\_info WHERE state\_name = `Virginia' ORDER BY year\_built ASC LIMIT 3;"
\newline Detected valid SQL, executing...
\newline[Coordinator] Routing sub-question: `Why do the maintenance costs of these bridges tend to be higher according to engineering guidelines?'
\newline [Complex to Single Query Router] Decision for IR\_AGENT
\newline [Coordinator] Synthesizing final answer...
 \\
\hline
Q5 & [Agent Server] Received complex query for coordinator: `List five of the oldest bridges in the United States still in use today, and briefly describe their historical significance.'
\newline [Coordinator] Decomposing user query...
\newline [Coordinator] Decomposed into: [`What are the five oldest bridges in the United States still in use today?', `What is the historical significance of each of the five oldest bridges in the United States?']
\newline [Coordinator] Routing sub-question: `What are the five oldest bridges in the United States still in use today?'
\newline [Complex to Single Query Router] Decision for SQL\_AGENT
\newline Raw LLM output: `SELECT structure\_number, year\_built FROM bridge\_basic\_info WHERE year\_built < 1970 ORDER BY year\_built ASC LIMIT 5;'
\newline Detected valid SQL, executing...
\newline [Coordinator] Routing sub-question: `What is the historical significance of each of the five oldest bridges in the United States?'
\newline [Complex to Single Query Router] Decision for GENERAL\_AGENT
\newline [Coordinator] Synthesizing final answer...
 \\
\hline
Q6 & [Agent Server] Received complex query for coordinator: `Compare the advantages and disadvantages of concrete arch bridges and steel truss bridges in terms of maintenance, lifespan, and load capacity' 
\newline [Coordinator] Decomposing user query...
\newline [Coordinator] Decomposed into: [`What are the advantages of concrete arch bridges in terms of maintenance?', `What are the disadvantages of concrete arch bridges in terms of maintenance?', `What is the lifespan of concrete arch 
bridges?', `What is the load capacity of concrete arch bridges?', `What are the advantages of steel truss bridges in terms of maintenance?', `What are the disadvantages of steel truss bridges in terms of maintenance?', `What is the lifespan of steel truss bridges?', `What is the load capacity of steel truss bridges?']
\newline [Coordinator] Routing sub-question: `What are the advantages of concrete arch bridges in terms of maintenance?'
\newline [Complex to Single Query Router] Decision for IR\_AGENT
\newline [Coordinator] Routing sub-question: `What are the disadvantages of concrete arch bridges in terms of maintenance?'
\newline [Complex to Single Query Router] Decision for IR\_AGENT
\newline [Coordinator] Routing sub-question: `What is the lifespan of concrete arch bridges?'
\newline [Complex to Single Query Router] Decision for IR\_AGENT
\newline [Coordinator] Routing sub-question: `What is the load capacity of concrete arch bridges?'
\newline [Complex to Single Query Router] Decision for IR\_AGENT
\newline [Coordinator] Routing sub-question: `What are the advantages of steel truss bridges in terms of maintenance?'
\newline [Complex to Single Query Router] Decision for IR\_AGENT
\newline [Coordinator] Routing sub-question: `What are the disadvantages of steel truss bridges in terms of maintenance?'
\newline [Complex to Single Query Router] Decision for IR\_AGENT
\newline [Coordinator] Routing sub-question: `What is the lifespan of steel truss bridges?'
\newline [Complex to Single Query Router] Decision for IR\_AGENT
\newline [Coordinator] Routing sub-question: `What is the load capacity of steel truss bridges?'
\newline [Complex to Single Query Router] Decision for IR\_AGENT
\newline [Coordinator] Synthesizing final answer...
 \\
\hline
\end{tabular}

\caption{Automated actions for complex queries (Q4-Q6) in the backend.}
\label{table: table_backend2}
\end{table*}

\end{document}